\documentclass{aa}  

\usepackage{graphicx}
\usepackage{tabularx} 
\usepackage{float}
\usepackage{txfonts}
\usepackage{orcidlink}
\usepackage{amsmath}
\usepackage{textcomp}
\begin{document}

   \title{Assessing methods for telluric removal on atmospheric retrievals of high-resolution optical exoplanetary transmission spectra}


   \author{Cathal Maguire \orcidlink{0000-0002-9061-780X}\inst{1}\thanks{E-mail:  \href{mailto:maguic10@tcd.ie}{maguic10@tcd.ie}}
          \and
          Elyar Sedaghati \orcidlink{0000-0002-7444-5315}\inst{2}
          \and
          Neale P. Gibson \orcidlink{0000-0002-9308-2353}\inst{1}
          \and
          Alain Smette \inst{2}
          \and
          Lorenzo Pino 
          \orcidlink{0000-0002-1321-8856}\inst{3}
          }

   \institute{School of Physics, Trinity College Dublin, University of Dublin, Dublin 2, Ireland
         \and
             European Southern Observatory, Alonso de Córdova 3107, Santiago, Chile
        \and
        INAF – Osservatorio Astrofisico di Arcetri, Largo Enrico Fermi 5, 50125 Firenze, Italy
             }

   \date{Received XXXX; accepted XXXX}

 
  \abstract
   {Recent advancements in ultra-stable ground-based high-resolution spectrographs have propelled ground-based astronomy to the forefront of exoplanet detection and characterisation. However, the resultant transmission and emission spectra of exoplanetary atmospheres are inevitably contaminated by telluric absorption and emission lines due to the light's transmission through the Earth's atmosphere above the observatory. Retrieving accurate atmospheric parameters depends on accurate modelling and removal of this telluric contamination while preserving the faint underlying exoplanet signal.}
   {There exist many methods to model telluric contamination, whether directly modelling the Earth's transmission spectrum via radiative transfer modelling, or using a principal component analysis (PCA)-like reconstruction to fit the time-invariant features of a spectrum, and removing these models from the observations. We aimed to assess the efficacy of these various telluric removal methods in preserving the underlying exoplanetary spectra.}
   {We compared two of the most common telluric modelling and removal methods, \textsc{molecfit} and the PCA-like algorithm \textsc{SysRem}, using planetary transmission spectra injected into three high-resolution optical observations taken with ESPRESSO. These planetary signals were injected at orbital periods of P = 2 days and P = 12 days, resulting in differing changes in radial velocity during transit. We then retrieved various injected atmospheric model parameters in order to determine the efficacy of the telluric removal methods, as well as their effect on the transmission spectra of exoplanets with varying orbital architectures.}
   {For the close-in, high velocity injected signal, we found that \textsc{SysRem} performed better for species that are also present in the Earth's atmosphere--with accurate and precise retrieval of atmospheric abundances and $T$-$P$ profiles, across each of the datasets. As we moved to slower moving signals at larger orbital separations, for one of the three datasets, \textsc{SysRem} dampened the planetary H$_2$O signal, leaving the retrieved abundance value unconstrained. In contrast, the H$_2$O signal was preserved for the telluric modelling method, \textsc{molecfit}. However, this behaviour was not ubiquitous across all three of the injected datasets, with another dataset showing a more precise H$_2$O/Fe ratio when preprocessed with \textsc{SysRem}. These conflicts highlight the importance of testing individual high-resolution dataset reduction routines independently to ensure real exoplanetary signals are preserved.}
   {}

\titlerunning{Assessing telluric removal methods on high-resolution exoplanetary transmission spectra}
\authorrunning{C. Maguire et al.}
\maketitle
%

\section{Introduction}
Recent advancements in ultra-stable ground-based high-resolution (R $\ge 25,000$) spectrographs have ushered in a new era of exoplanetary characterisation, via transmission and emission spectroscopy studies. These observations allow for the unambiguous detection of atmospheric atomic and molecular species, which can be used to infer atmospheric dynamics, such as winds and rotation \citep{Snellen_2010,Brogi_2012,Seidel_2020}, as well as potentially map longitudinal variations in the abundance profiles of atmospheric species \citep{Ehrenreich_2020,Kesseli_2021,Prinoth_2022,Gandhi_2022,Gandhi_2023,Maguire_2024}. They have also been developed upon to constrain the (relative) abundances of these species \citep{Gandhi_2019,Line_2021,Pelletier_2021,Gibson_2022,Maguire_2023} and ultimately link a planet's current composition to its formation and evolution pathway. 

The carbon-to-oxygen (C/O) ratio measured in an exoplanetary atmosphere has long been thought to grant an insight into where in the protoplanetary disk a planet formed and how it migrated to its current orbital distance \citep{Oberg_2011,Madhusudhan_2014,Mordasini_2016}. \cite{Lothringer_2021} also showed that the relative abundance of neutral metals, or refractories, to that of volatile species such as H$_2$O and CO, in the form of a ``refractory-to-volatile'' ratio, could be used to constrain the formation and migration pathway of hot/ultra-hot Jupiters (HJ/UHJs). It is, however, worth noting that this process is more complex than simply inferring a planet's formation pathway directly from its current composition, due to the numerous complex processes which impact a planet’s formation and evolutionary history \citep{vanderMarel_2021,Molliere_2022}. Furthermore, oxygen abundances are difficult to estimate from detected species alone, as oxygen-bearing species are subject to thermal dissociation in UHJs, and/or condensation and subsequently rainout in the form of enstatite and forsterite at cooler temperatures \citep{Burrows_2001}.  Despite this, abundance ratios measured from the ground--whether C/O, Fe/H, refractory-to-volatile, or isotope ratios--may be used in tandem to infer where these objects formed and how they evolved with time. Recent work by \cite{Line_2021} constrained the C/O ratio of the hot-Jupiter WASP-77Ab to within $\sim$~0.1 dex, using the Immersion GRating INfrared Spectrometer (IGRINS) at the Gemini South Observatory, highlighting the enormous importance and potential of ground-based high-resolution spectroscopy alongside low-resolution space-based observations with the James Webb Space Telescope (JWST) or the Hubble Space Telescope (HST). 

However, observing from the ground poses significant challenges, not least the removal of contamination by numerous telluric absorption and emission lines caused by the Earth’s atmosphere. In the past, a telluric standard star would be observed alongside the target star, in order to correct for telluric contamination. This method, however, is costly, as it would require valuable telescope time to observe the telluric star, which would otherwise be used for other scientific observations \citep{Smette_2015}. This method of telluric correction is also not perfect, as temporal variation of the Earth's atmosphere over the course of an observation causes differences in the contamination of the reference and target spectra, particularly if both are not subsequently observed within a short time period and/or at small angular separations on the sky. Furthermore, normalising a telluric standard's continuum, as well as its broad absorption lines, is never without imprecisions. The contamination from telluric lines is significantly worse in the near-infrared/infrared (NIR/IR) where molecular absorption dominates the spectrum. At these wavelengths, telluric contamination--particularly from H$_2$O--can often saturate the absorption spectrum. Therefore, the proper modelling and removal of these telluric bands are of the utmost importance in the search for telluric signatures in exoplanetary atmospheres from the ground.

In more recent years, it has become common practice to model the telluric spectra directly, with tools such as \textsc{molecfit} \citep{Smette_2015,Kausch_2015} and \textsc{TelFit} \citep{Gullikson_2014}, which combine line-by-line radiative transfer calculations with on-site meteorological data to estimate the transmission spectrum of the Earth's atmosphere across the wavelength range of the observations. This method also accounts for the temporal variation of airmass above the telescope over the course of an observation, and has been successfully implemented in removing telluric contamination of high-resolution exoplanetary spectra \citep{Sedaghati_2021}. This telluric removal method has also been shown to be more effective in removing the contamination on high-resolution X-Shooter spectra compared to correcting with a telluric reference star \citep{Kausch_2015}. It has also been shown to be more effective in removing the contamination on exoplanetary spectra when compared to obtaining a telluric reference spectrum directly from the data, assuming telluric line strength varies linearly with airmass \citep{Langeveld_2021}.

Another common method of telluric removal is constructing a model of the observed spectra directly using principal component analysis (PCA) algorithms and removing this model from the observations. Many forms of PCA have been successfully implemented to treat high-resolution spectroscopy observation of exoplanetary atmospheres, such as singular value decomposition \citep{de_Kok_2013}, eigenvalue decomposition \citep{Damiano_2019}, and \textsc{SysRem} \citep{Tamuz_2005, Birkby_2013}. In contrast to PCA, for example, \textsc{SysRem} can account for time- and wavelength-based uncertainties when modelling systematic trends. The modelling and removal of telluric contamination from high-resolution exoplanetary spectra using \textsc{SysRem} was first implemented by \cite{Birkby_2013}, and has since become common practice for both transmission and emission observations \citep[e.g.,][]{Gibson_2020,Nugroho_2021}. These methods model the stellar and telluric lines which are (quasi-)stationary in wavelength, as a function of time, which can then be subsequently removed from the observations. However, these algorithms can often be overly aggressive in their removal, as there exists no obvious halting criterion, and thus run the risk of inadvertently distorting the underlying exoplanet signal \citep{Birkby_2017}. There are also other emerging techniques to model time-varying systematics on high-resolution spectra which are worth noting.

\cite{Meech_2022} modelled the telluric and planetary components of high-resolution dayside emission spectra using Gaussian process (GP) regression, and successfully recovered injected planetary signals. They also showed that traditional telluric removal frameworks, such as linear airmass regression and \textsc{SysRem}, significantly truncate planetary atmospheric features in comparison to their GP methodology. Recent work by \cite{Cabot_2024} and \cite{Cheverall_2024} successfully recovered a simulated signal of the temperate sub-Neptune TOI-732c from IGRINS data using PCA-based telluric removal, showing the feasibility of these detrending methods at low radial velocities changes during transit. \cite{Cheverall_2024} conclude that due to the high number of out-of-transit spectra observed, the planetary signal does not form a principal component of the spectra, and thus the planetary signal is prevented from being removed by the detrending algorithm, albeit it can be significantly altered by it.

When performing model comparison, in the form of an atmospheric retrieval or otherwise, it is imperative that the inevitable distortion on the underlying exoplanetary signal is taken into account, such that the forward model is an accurate representation of the noiseless data. If the distortions caused by these detrending algorithms are not accounted for, they may bias the retrieved atmospheric parameters \citep{Brogi_Line_2019}. Previous studies have addressed this issue by filtering the forward model in the same way as the data \citep{Brogi_Line_2019,Pelletier_2021}. However, due to the need to interpolate the 1D model to the 3D data array (order $\times$ time/phase $\times$ wavelength) and then reprocess each forward model for each likelihood calculation in an atmospheric retrieval framework, these methods generally require a large amount of computing resources. \cite{Gibson_2022} recently implemented a fast model filtering technique, which uses the \textsc{SysRem} model basis vector(s) to reproduce these distortions on the atmospheric forward model. This technique can also be applied where standard PCA algorithms have been used to detrend the data.

The precise removal of telluric contamination of ground-based spectra, whilst also preserving the faint underlying exoplanet spectra, is of the utmost importance to the exoplanet community. This will allow for the optimal extraction of the information-rich datasets on the horizon.

The study is summarised as follows. In Sect.~\ref{sec:Method} we present the three ESPRESSO observations into which we inject two planetary models of P = 2 days and P = 12 days, respectively. We then proceed to outline both telluric modelling/removal methods. In Sect.~\ref{sec:Analysis} we demonstrate the recovery of the injected signals, including traditional cross-correlation analysis as well as atmospheric retrievals of both injection scenarios on each of the three datasets. In Sect.~\ref{sec:Discussion} we discuss the findings before concluding the study and summarising the results in Sect.~\ref{sec:Conclusions}.

\section{Methods}
\label{sec:Method}
\subsection{ESPRESSO observations}
\label{sec:ESP_Obs}
The injection and recovery tests detailed henceforth were performed on three time-series observations of the UHJ host MASCARA-4 \citep{Dorval_2020} with the Echelle SPectrograph for Rocky Exoplanets and Stable Spectroscopic Observations (ESPRESSO) installed at the Very Large Telescope (VLT), Paranal Observatory, Chile \citep{Pepe_2021}. These observations were taken on the nights of 2022 February 4, 2022 February 6, and 2022 February 9 as part of Program ID 106.21BV.005 (PI: Pino). These datasets are hereafter referred to as DS1, DS2, and DS3, respectively. These datasets were chosen in order to simulate realistic time-series observations of an UHJ host star. These observations were performed for the purpose of measuring the planet's emission spectrum, and are therefore free of any transit signatures in the high-resolution spectra. These signatures typically include spectral artefacts on the in-transit stellar lines due to the inhomogeneity of the stellar disc as the planet transits, caused by the centre-to-limb variation (CLV) and the Rossiter-McLaughlin (RM) effect \citep{Rossiter1924,McLaughlin1924}. This choice was made to minimise the post-processing steps, focusing solely on the effects of telluric removal methods. As they are emission spectroscopy observations, they also minimise contamination from a potentially transiting exoplanet's signal, as found in similar time-series observations. MASCARA-4 is an A-type star, thus these datasets were chosen as the star has very broad absorption lines (see Figs.~\ref{molecfit} \&~\ref{Prep_steps}), allowing for precise modelling of telluric lines. The data were reduced using the ESPRESSO Data Reduction Software (DRS) v2.4.0, with the extracted non-blaze-corrected S2D spectra ({\tt ESO PRO CATG = S2D\_BLAZE\_A}) being used for further analysis. It must be noted that we made the decision not to use the sky-subtracted spectra, to avoid adding unnecessary noise into the spectra. However, this has the disadvantage of not removing the few sky emission lines (mainly from O$_2$ and OH) present in the wavelength range of ESPRESSO. This, however, has implications for the retrieved relative abundances of OH, which is discussed further in Sect.~\ref{sec:AtmoRetDiscuss_P2}.
\subsection{Forward model}
\label{sec:ForMod}
For the injected signal, we assume two scenarios; a Jupiter mass planet with an orbital period of 2 and 12 days, hereafter referred to as P2 and P12, respectively. The stellar and planetary parameters of the injected models are given in Table~\ref{tab:Table1}.
For the injected forward models we generated 1D transmission spectra, using the \textsc{irradiator} code, described in detail in \cite{Gibson_2022}, computed across a wavelength range of 3670$-$8000$\AA$, at a constant resolution of R $=$ 200,000. We included multiple atmospheric opacity sources, both molecular and atomic, with significant absorption in the given wavelength range. These included Fe, V, Na, OH, and H$_2$O. The cross-sections for Fe and V were obtained via the Kurucz line list \citep{Kurucz_1995}. Similarly, the Na cross-sections were obtained via the VALD line list \citep{Piskunov_1995}. The OH and H$_2$O cross-sections were obtained via the MoLLIST \citep{Bernath_2020} and POKAZATEL \citep{Polyansky_2018} line lists, respectively. These opacity sources are assumed to have a constant volume mixing ratio (VMR) with altitude, whilst assuming a vertical temperature-pressure ($T$-$P$) profile parameterised using the treatment outlined in \cite{Guillot_2010}. This treatment requires four distinct parameters: $T_{\rm irr}$, $T_{\rm int}$, $\log_{10}(\kappa_{\rm IR})$, $\log_{10}(\gamma)$, corresponding to the irradiation temperature, internal temperature, mean infrared opacity, and the ratio of visible-to-infrared opacity, respectively. We also included a H$_2$ VMR, $\chi_{\rm ray}$, along with H$_2$ opacities from \cite{Dalgarno_1962}, to account for Rayleigh scattering, as well as a further continuum opacity source caused by an opaque cloud deck at a pressure level $P_{\rm cloud}$, below which the transmission spectrum model is truncated. Thus, for the one-dimensional atmospheric model, we have 6 $+$ N$_{\rm species}$ input parameters ($T_{\rm irr}$, $T_{\rm int}$, $\log_{10}(\kappa_{\rm IR})$, $\log_{10}(\gamma)$, $\log_{10}(P_{\rm cloud})$, $\log_{10}(\chi_{\rm ray})$, $\log_{10}(\chi_{\rm species})$). The transmission spectra for each injection scenario are shown in Fig.~\ref{Injected_models}. For each injection scenario we calculate the planet's irradiation temperature, $T_{\rm irr}$, from Eq.~1 of \cite{Guillot_2010}, assuming a circular orbit. It is worth noting that at the larger orbital distances, the composition of this class of planet would differ significantly from that of a UHJ at P = 2 days. Although the decrease in temperature is accounted for in the injected models, the underlying chemical composition is not representative of a planet at this orbital distance. As we aim to evaluate biases introduced by the various telluric removal routines under varying planetary orbital architectures---particularly on time invariant signals---as opposed to biases introduced by varying chemistry, the chemical composition is kept constant. These transmission spectra were then convolved with a number of convolution kernels to imitate the astrophysical and instrumental effects which change the shapes of the final observed spectral lines. Firstly, the spectra were convolved with a box kernel whose width was determined from the maximum velocity shift between subsequent exposures over the course of the observations to simulate signal `smearing'. For P2, the overall change in planetary radial velocity during transit was $\sim 19$~km~s$^{-1}$, whereas for P12, this change was $\sim 10$~km~s$^{-1}$.  Secondly, the spectra were convolved with the mean instrumental profile of ESPRESSO, the full-width at half-maximum (FWHM) of which was determined by fitting Gaussian profiles to the spectrograph's calibration source ThAr lines across the full wavelength range of ESPRESSO. Finally, to account for solid-body rotational broadening, the spectra were convolved with another Gaussian kernel with a FWHM equal to the planetary rotational velocity, $v_{\rm rot}$, assuming a tidally-locked planet. The final assumption of a Gaussian rotational broadening kernel is not entirely accurate, as a rotating atmosphere is not a sphere but an annulus with a distinct rotational broadening kernel, which has been studied extensively in recent works \citep[e.g.,][]{Gandhi_2022,Gandhi_2023,Boucher_2023,Maguire_2024,Nortmann_2024}. However, for the purposes of this study, a Gaussian rotational kernel is preferred for simplicity. It is worth noting that we do not retrieve the rotational velocity of the planet, however the line shapes and the resultant retrieved broadening may be subject to distortions caused by the various preprocessing methods outlined in Sects.~\ref{sec:molecfit} \& \ref{sec:sysrem}. After this point, the signals were injected into the observed data in slightly different manners, depending on the final treatment of the tellurics, outlined below.
\subsection{Model injection}
\label{sec:ModInj}
For the \textsc{molecfit} procedure, for each orbital phase, $\boldsymbol{\phi}$, the 1D convolved transmission spectra were Doppler shifted by the injected planetary velocity given by:
\begin{align}
 \boldsymbol{v}_{\rm \textbf{p}} &= K_{\rm p}\sin{(2\pi\cdot \boldsymbol{\phi} )} + \Delta v_{\rm sys} + \boldsymbol{v_{\rm sys}} 
 \label{eq:planet_vel}
\end{align}
where $K_{\rm p}$, $\boldsymbol{\phi}$, and $\Delta v_{\rm sys}$ are the injected planet's radial velocity semi-amplitude, orbital phase, and systemic velocity offset, respectively. $\boldsymbol{v_{\rm sys}}$ is the systemic velocity, which is corrected for in Sects.~\ref{sec:molecfit} \&~\ref{sec:sysrem}. This 2D grid (time/phase $\times$ wavelength) is then directly injected into the raw, merged S1D ESPRESSO spectra, which \textsc{molecfit} uses to compute the telluric model. Injecting into the S1D spectra is a necessary additional step for the \textsc{molecfit} preprocessing since this is the data product \textsc{molecfit} utilises. Therefore, to replicate any effects \textsc{molecfit} may impose on a planetary signal, the model must also be injected into the S1D data product before running \textsc{molecfit}.
For both the \textsc{molecfit} and \textsc{SysRem} procedures, the 1D convolved transmission spectra were also linearly interpolated to the 2D wavelength grid of the data (order $\times$ wavelength). For each order, the 2D forward models were Doppler shifted to the injected planetary velocity given by Eq.~\ref{eq:planet_vel}. This 3D (order $\times$ time/phase $\times$ wavelength) array was then injected into the raw S2D ESPRESSO spectra. These are the final datasets which undergo the respective telluric removal procedures, outlined below.
\begin{table}
	\centering
	\caption{Injected stellar and planetary parameters}
	\label{tab:Table1}
	\begin{tabular}{cll} 
		\hline
		\hline
				Parameter & P2 & P12\\
		\hline
		\hline
    \multicolumn{1}{c}{} & \multicolumn{1}{c}{\hspace{1.0cm}}\\[-0.2cm]
    $R_\star$ & 1.0 R$_{\odot}$& 1.0 R$_{\odot}$\\[0.1cm]
    $M_\star$ & 1.0 M$_{\odot}$& 1.0 M$_{\odot}$\\[0.1cm]
    $T_{\rm eff}$ & 5780 K & 5780 K\\[0.1cm]
    $P$ & 2.0 days & 12.0 days\\[0.1cm]
    $R_{\rm p}$ & 1.25 R$_{\rm Jup}$& 1.25 R$_{\rm Jup}$\\[0.1cm]
    $M_{\rm p}$ & 0.75 M$_{\rm Jup}$& 0.75 M$_{\rm Jup}$\\[0.1cm]
    $a_{\rm p}$/$R_\star$ & 6.68 & 22.06\\[0.1cm]
    $K_{\rm p}$ &  -104.10 km s$^{-1}$ &  -57.29 km s$^{-1}$ \\[0.1cm]
    $T_{\rm irr}$ & 2236.31 K & 1230.69 K \\[0.1cm]
    $R_{\rm p}$/$R_\star$ & 0.1541 & 0.1541 \\[0.1cm]
    $e$ & 0.0 & 0.0 \\[0.1cm]
    $i$ & 0.0$^{\circ}$ & 0.0$^{\circ}$ \\[0.1cm]
    \hline
    \end{tabular}

\end{table}
\subsection{Spectral preprocessing with \textsc{molecfit}}\label{sec:molecfit}
\begin{figure}
    \includegraphics[width=\columnwidth]{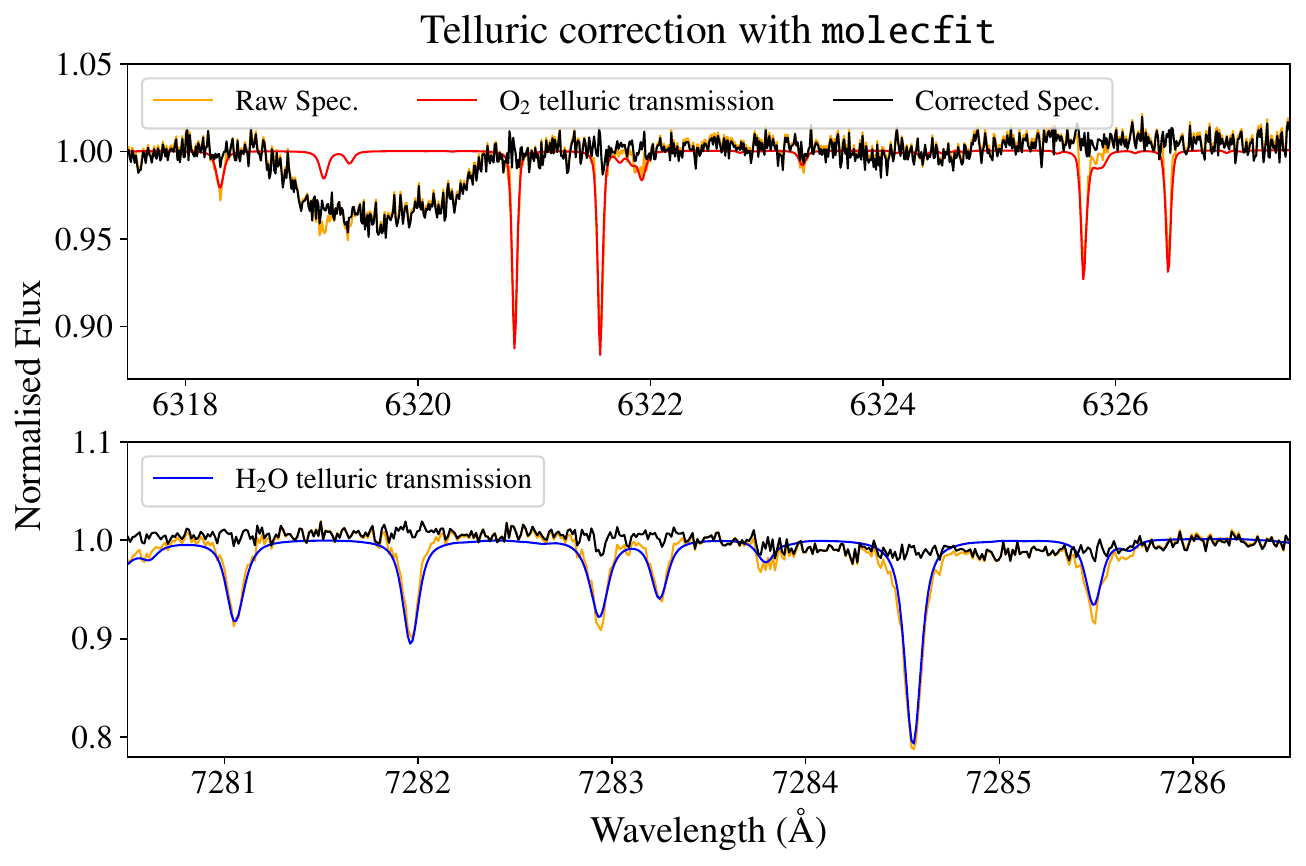}
    \caption{Example of telluric absorption correction with \textsc{molecfit} on ESPRESSO S1D spectra. The top panel shows an O$_2$ region, including a broad stellar absorption line. The bottom panel shows a H$_2$O region. The final telluric model was used to correct the raw ESPRESSO S2D spectra (see Sect.~\ref{sec:molecfit})} 
    \label{molecfit}
\end{figure}
To remove the telluric absorption signature with \textsc{molecfit}, the S1D (stitched and resampled) spectra from the pipeline were used. The \textsc{molecfit} (v. 4.2.3) routines were run on esoreflex platform independently for each spectrum. Telluric absorption lines were modelled with a line by line radiative transfer model, using the atmospheric pressure profile that \textsc{molecfit} estimated from the temperature and humidity profiles that it obtains from the Global Data Assimilation System (GDAS) database. These profiles were used as starting point in a minimisation algorithm that finds the best fit model to the telluric absorption lines for those molecules input by the user. In the case of ESPRESSO's wavelength coverage, only O$_2$ and H$_2$O were included in the modelling. In this modelling procedure, we fit for the parameters of a variable kernel (a Voigt profile) describing line shapes, a first order polynomial describing the local continuum at the location of the selected absorption lines, and the abundances of the two selected species. From the results produced by \textsc{molecfit}, the normalised telluric transmission model was used to remove the telluric contamination from the spectra, details of which are given below. An example of this procedure is shown in Fig.~\ref{molecfit}. Future processing by \textsc{molecfit} would likely gain by using the updated ESPRESSO instrumental Line Spread Function (LSF) determined by \cite{Schmidt_2024}.

Apart from applying the respective telluric correction routines, \textsc{molecfit} or \textsc{SysRem} (see Sect.~\ref{sec:sysrem} below), the data reduction employs identical steps:
\begin{enumerate}[]
    \item[\textbf{(1)}]The time- and wavelength-dependent uncertainties of the raw S2D data are estimated assuming Poisson noise, following the procedure outlined in \cite{Gibson_2020}.
    \item[\textbf{(2)}]We generate a telluric model of the S1D (time/phase $\times$ wavelength) data using v4.2.3 of \textsc{molecfit}. This model is then linearly interpolated to the S2D grid (order $\times$ time/phase $\times$ wavelength), before dividing it out of the model-injected S2D data.
    \item[\textbf{(3)}]The paired \'{e}chelle slices, being the result of ESPRESSO's Anamorphic Pupil Slicing Unit (APSU), are merged into single orders.
    \item[\textbf{(4)}]These merged spectra are then shifted to the stellar rest frame using the ESPRESSO pipeline radial velocities, $\boldsymbol{v_{\rm sys}}$.
    \item[\textbf{(5)}] The orders are placed on a common blaze function, by firstly dividing each order by its mean spectrum. The resultant residual array was then smoothed with a median filter to remove outliers,
    followed by a Gaussian filter. The original orders and their respective uncertainties were then divided by these resultant smoothed residuals. This procedure places each spectrum on a common blaze function in time.
    \item[\textbf{(6)}] Each order is divided by a master stellar spectrum, which is created using the mean of the out-of-transit spectra in each order.

\end{enumerate}
\begin{figure*}[!htbp]
    \centering
    \includegraphics[width=\textwidth]{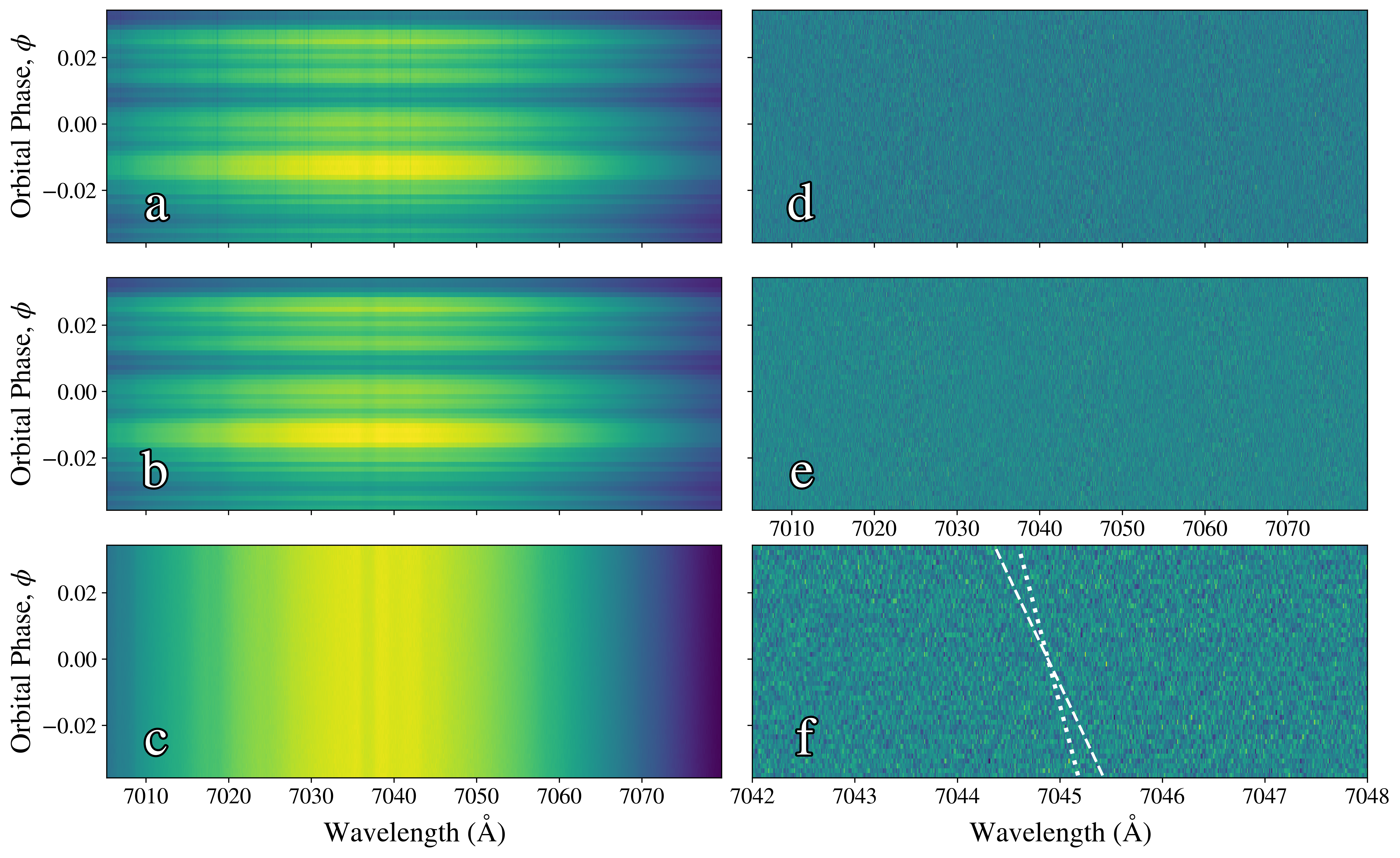}
        \caption{Various reduction steps applied for an individual ESPRESSO order of DS1. \textbf{(a)} Before telluric removal with \textsc{molecfit}. \textbf{(b)} After telluric removal with \textsc{molecfit} and shifted to the stellar rest frame. \textbf{(c)} Each spectrum is placed on a common blaze function (see Sect.~\ref{sec:molecfit}). \textbf{(d)} Residuals after division through by the mean out-of-transit spectrum, weighted on their uncertainties. This panel shows the final data product used in the \textsc{molecfit} preprocessing analysis. \textbf{(e)} Residuals after division through by the mean out-of-transit spectrum, weighted on their uncertainties, and subtraction of the \textsc{SysRem} model. This panel shows the final data product used in the \textsc{SysRem} preprocessing analysis. It is worth noting that for this procedure, we do not perform any preceding telluric removal (i.e., Sect.~\ref{sec:molecfit}, Step (2)). \textbf{(f)} A magnification of panel (e) showing the wavelength shifts caused by each injected model. The planetary velocity curve for P2 is shown as a white dashed curve, whereas the slower moving planet's (P12) velocity curve is shown as a white dotted curve.} 
        \label{Prep_steps}
\end{figure*}
\subsection{Spectral preprocessing with \textsc{SysRem}}
\label{sec:sysrem}
Steps \textbf{(1)}, and \textbf{(3)}-\textbf{(6)} are performed identically for the \textsc{SysRem} correction. For each \textsc{SysRem} iteration, each order is decomposed into two column vectors, \textbf{u} and \textbf{w}, where the \textsc{SysRem} model for each order is given by their outer product, \textbf{uw}$^{\rm T}$. This model is then subtracted from the input array(s), to get the processed array(s), and the procedure is repeated on the processed array(s) for the next iteration. Dividing through by the median spectrum in each order first and then subsequently subtracting the \textsc{SysRem} model allows for a faster model filtering process in each likelihood calculation \citep{Gibson_2022}. It is worth noting that a more accurate approach would be to perform \textsc{SysRem} on the unnormalised data, and divide by the resultant model, however we expect this difference to be negligible here. The effects of each of the preprocessing steps outlined above on a single ESPRESSO order are highlighted in Fig.~\ref{Prep_steps}. 
For each of the datasets, and each of the injected models, the number of \textsc{SysRem} iterations was optimised following the procedure outlined in Sect.~2.3 of \cite{Maguire_2023}. The optimum number of \textsc{SysRem} iterations for each dataset was determined from the resultant detection significance of the injected model, using the likelihood distribution of the model scale factor, $\alpha$, conditioned on the injected $K_{\rm p}$, and $\Delta v_{\rm sys}$ values. The expected value, or mean, of this distribution was computed, and divided by its standard deviation, resulting in a detection significance. This, in essence, computes the number of standard deviations the maximum signal deviates from an $\alpha$ of 0 (i.e., a non-detection). For DS1, we found the optimum number of iterations to be 13 and 10 for P2 and P12, respectively. Likewise, for DS2 the optimum number of iterations were found to be 3 and 1 for P2 and P12, respectively. Finally, for DS3, the optimum number of iterations were found to be 8 and 8 for P2 and P12, respectively. Although there is no consensus on an optimum stopping criterion for PCA-like telluric removal methods, this method is preferred here as it does not suffer from biases introduced by noise amplification, as we are determining the detection significance from the likelihood as a function of alpha directly, conditioned on the injected $K_{\rm p}$, and $\Delta v_{\rm sys}$. These biases are seen for detection significances computed directly from cross-correlation functions (CCF) and their resultant $K_{\rm p}$--$\Delta v_{\rm sys}$ maps \citep{Cheverall_2023}. The problem of determining a stopping criterion at which the `optimum' number of iterations is reached via injection-recovery tests is completely model-dependent. Furthermore, in principle this optimum number would change order-by-order depending on the severity of the telluric contamination, as well as additional systematics, present in each order.
\section{Analysis}
\label{sec:Analysis}
\subsection{Injection and recovery tests}
\label{sec:InjTests}
\begin{figure*}[!htbp]
    \centering
    \includegraphics[width=\textwidth]{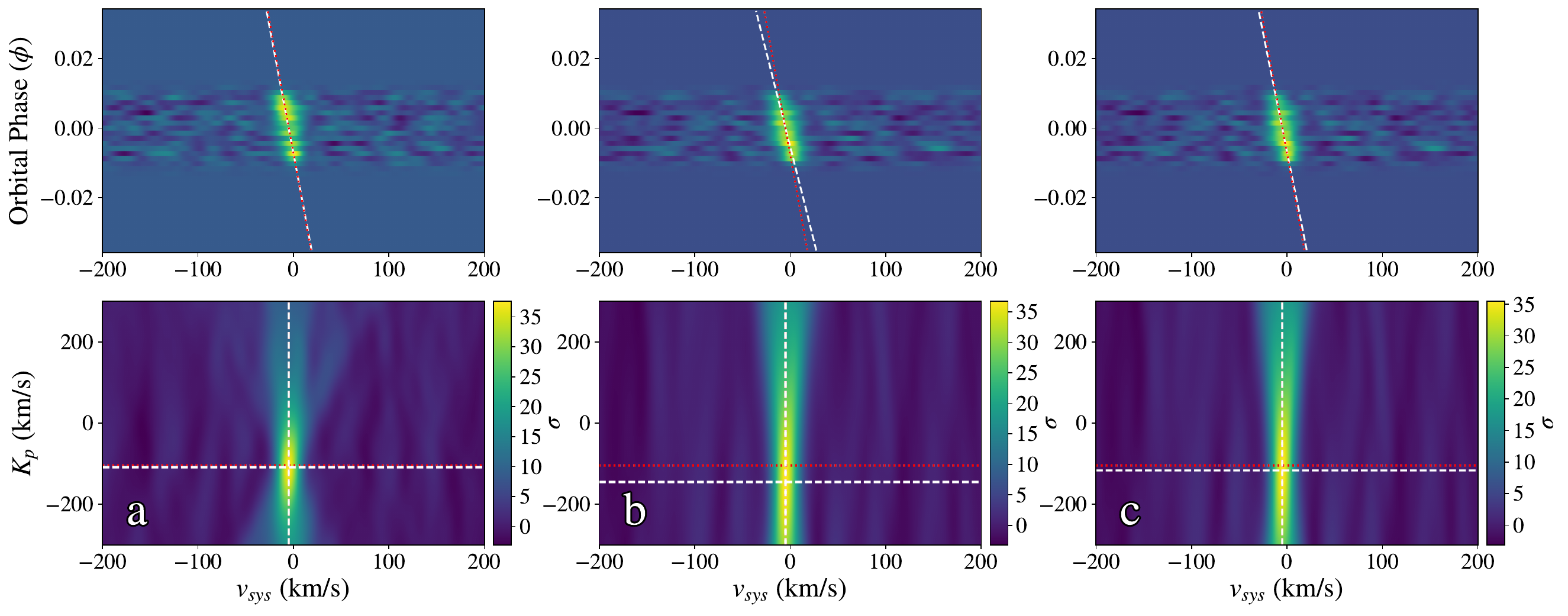}
        \caption{Cross-correlation and $K_{\rm p}$--$\Delta v_{\rm sys}$ maps for a planetary signal injected with P = 2 days injected into DS1 (P2). Red dotted lines/curves show the injected velocity values, whereas white dashed lines/curves show the recovered values (i.e., the values at which there was a maximum in the CCF). \textbf{(a)} Data reduced using the \textsc{SysRem} telluric removal method. Forward model also reprocessed using \textsc{SysRem} filtering technique \citep{Gibson_2022}. \textbf{(b)} Data reduced using the \textsc{molecfit} telluric removal method. Forward model also reprocessed by division of mean out-of-transit spectrum (see panel \textbf{(c)}). If this step was omitted, we found an anomalous offset from the injected $K_{\rm p}$ value.} 
        \label{Maps}
\end{figure*}
In the past, the efficacy of the removal of telluric contamination on high-resolution exoplanetary absorption/emission spectra has been determined in a variety of manners. A typical approach performs classic cross-correlation analysis with the injected forward model, and attempts to determine the telluric removal process which results in the highest `detection significance' of the injected signal \citep{Cheverall_2023}. Typically, in order to calculate the detection significance of a given model within a $K_{\rm p}$--$\Delta v_{\rm sys}$ map, the mean of the noise in the map in regions away from the peak is subtracted, before dividing the map by the standard deviation of the noise away from the peak \citep{Brogi_2012,Brogi_2018}. However, the regions away from the peak in which the mean/standard deviation of the noise is calculated are arbitrarily chosen, whilst also assuming the noise in these regions follow a Gaussian distribution (which is not always the case), and thus the resultant detection significance is not exact. Another method is to measure the line profiles of the atmospheric absorption/emission lines, and compare with the injected values \citep{Meech_2022} or compare the consistency of the atmospheric line profiles over the course of multiple observations \citep{Langeveld_2021}. 
With the recent introduction of Bayesian inference frameworks to high-resolution exoplanetary spectroscopy studies \citep{Brogi_Line_2019, Gibson_2020}, in the form of CC-to-likelihood mapping, we can directly compare the forward model to the data, after the aforementioned reduction steps, thus allowing for direct atmospheric retrievals of the injected signal. These `injection and recovery' tests not only allow us to test the efficacy of the telluric removal routines, but also can highlight any biases which may be introduced as a result of the various reduction procedures. This method also has the advantage of effectively averaging over the line profiles of multiple injected species, in order to assess any distortions caused by the telluric removal processes across the full wavelength range of the observations.
\subsection{Cross-correlation}
\label{sec:CC}
Traditional cross-correlation analysis \citep{Snellen_2010} was initially performed in order to recover the injected signal after both reduction procedures. This involves Doppler shifting an atmospheric model across a range of systemic velocity offsets, $\Delta v_{\rm sys}$, and cross-correlating with each of the time-series exposures. This effectively creates a two-dimensional time/orbital phase vs $\Delta v_{\rm sys}$ map, referred to as a “$\phi$--$\Delta v_{\rm sys}$” map, or simply, a “cross-correlation” map. This map will display the planetary cross-correlation trail projected to the injected orbital velocity given by Eq.~\ref{eq:planet_vel} with parameters outlined in Table~\ref{tab:Table1}. This cross-correlation map can also be shifted across a range of orbital velocities with varying $K_{\rm p}$ values and collapsed in time. This creates a map of cross-correlation values for each combination of $K_{\rm p}$ and $\Delta v_{\rm sys}$, known as a $K_{\rm p}$--$\Delta v_{\rm sys}$ map, which should reach a maximum at the injected values.
\subsection{Model filtering}
\label{sec:ModRP}
In order to accurately retrieve the injected model, the forward model with which one performs cross-correlation or likelihood analyses must be a representation of the noiseless data \citep{Gibson_2022}. Therefore, any distortions caused by the aforementioned reduction methods must be accounted for in the computation of the forward model. In the \textsc{SysRem}/PCA case, this includes performing a full \textsc{SysRem}/PCA filtering to the forward model, as outlined in \citep{Gibson_2022}. Whereas, in the \textsc{molecfit} preprocessing case, this means simply dividing the forward model by a master spectrum (ignoring the initial division of the telluric model). We find it is imperative, even simply for traditional cross-correlation analysis, to employ these model filtering methods, in order to recover the injected signal accurately.
\subsection{Atmospheric retrieval}
\label{sec:AtmoRet}
In order to recover the injected signal via an atmospheric retrieval analysis on each of the datasets, we must compute a forward model, $m_i$, for a given set of model parameters, $\boldsymbol{\theta}$. Assuming uniform priors, the log-posterior is computed by adding the log-prior to the log-likelihood:
\begin{align}
    \ln{\mathcal{L}} = - N\ln{\beta} - \frac{1}{2}\chi^2
    \label{eq:finL}
\end{align}
given
\begin{align}
        \chi^2 = \sum_{i=1}^{N}\frac{(f_i - \alpha m_i(\boldsymbol{\theta}))^2}{(\beta\sigma_i)^2}
        \label{eq:chi2}
\end{align}
\noindent where $f_i$ is the data, $\sigma_i^2$ is the data variance, $\alpha$ is the model scale factor, $\beta$ is the noise scale factor, and $N$ is the number of data points. See \cite{Gibson_2020} for a detailed derivation. We sampled the parameter space using a Markov Chain Monte Carlo algorithm (MCMC), specifically a Differential-Evolution Markov Chain (DEMC) \citep{TerBraak_2006,Eastman_2013}, running an MCMC chain with 128 walkers,  with a burn-in length of 500 and a chain length of 1000. We assessed each chains' convergence via the Gelman-Rubin (GR) statistic, after splitting each chain into four independent subchains. We also split the chains into groups of independent walkers, and overplot their 1D and 2D marginal distributions, in order to visualise the convergence of the MCMC. We compare the retrieved abundances of each of the injected species relative to the abundance of Fe, as the relative abundance constraints are typically more precise than those of the absolute abundances. This is due to a well-known degeneracy between the absolute abundances and other model parameters, such as the cloud deck pressure \citep{Benneke_Seager_2012,Heng_2017}. Due to the reduction methods, one loses the planetary continuum, thus we are only sensitive to the relative line strengths above the planet's continuum. This allows us to constrain the relative abundances more precisely, assuming we've accounted for the reduction methods when we compute the forward model (See Sect.~\ref{sec:ModRP}).
\begin{figure*}[!htbp]
    \centering
    \includegraphics[width=0.8\textwidth]{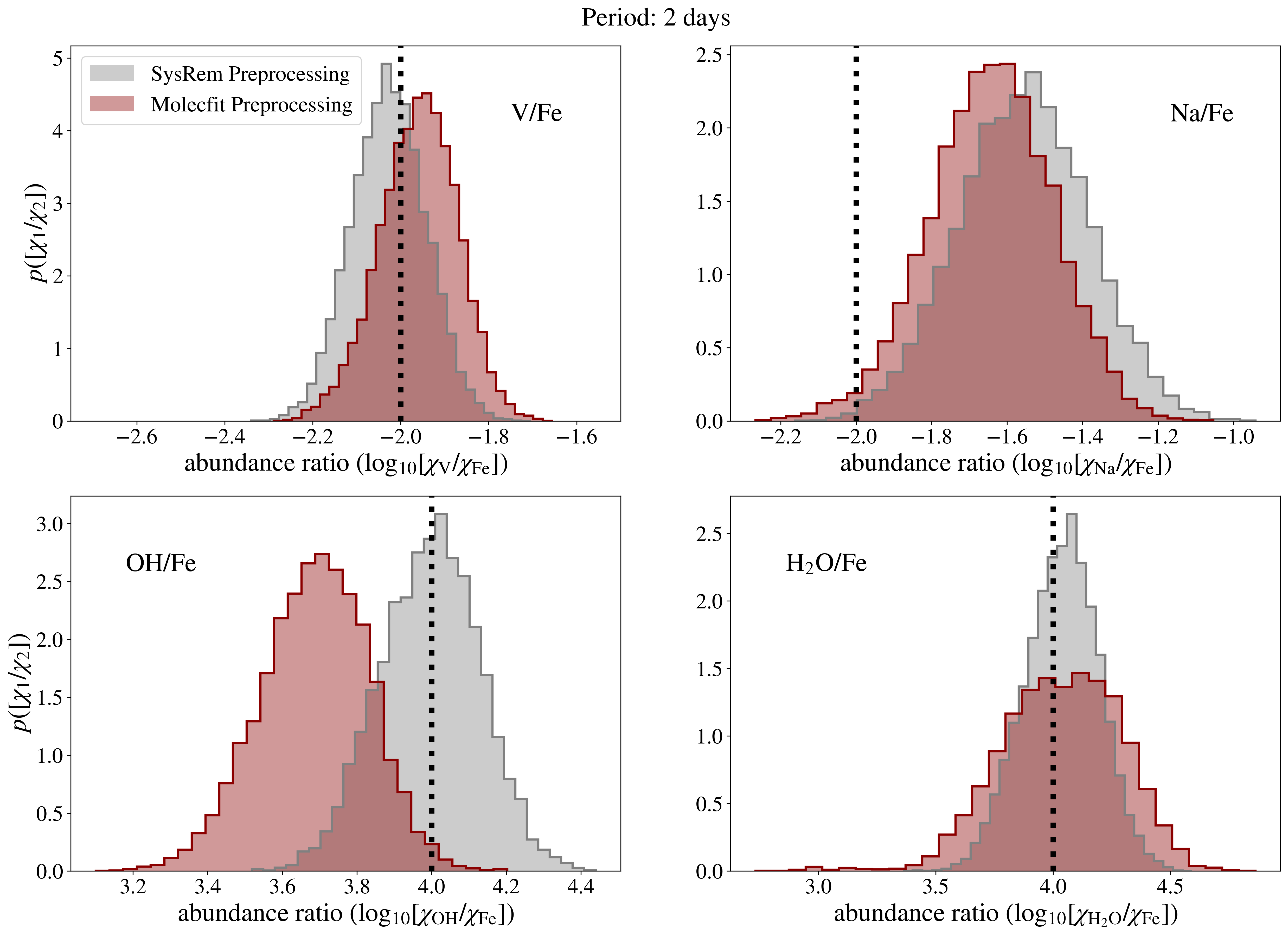}
        \caption{Relative abundance comparison for DS1, for P2. Injected relative abundance values shown as a vertical black dotted line. The posterior distributions for both the \textsc{SysRem} and \textsc{molecfit} preprocessing methods are shown in grey and red, respectively.} 
        \label{Obs1_P2_relabuns}
\end{figure*}
\begin{figure*}[!htbp]
    \centering
    \includegraphics[width=0.8\textwidth]{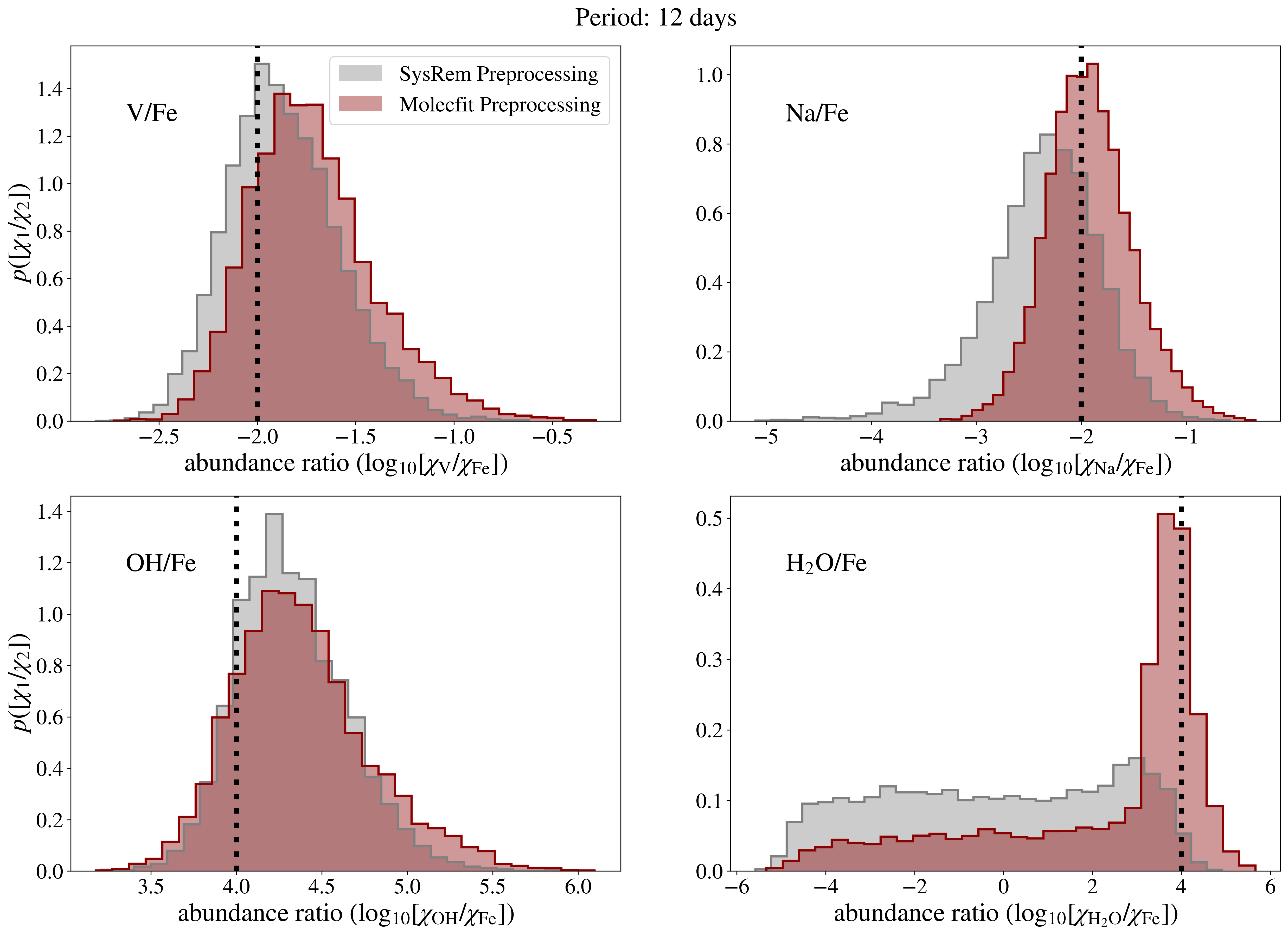}
        \caption{Relative abundance comparison for DS1, for P12. Injected relative abundance values shown as a vertical black dotted line. The posterior distributions for both the \textsc{SysRem} and \textsc{molecfit} preprocessing methods are shown in grey and red, respectively.} 
        \label{Obs1_P12_relabuns}
\end{figure*}
\section{Discussion}
\label{sec:Discussion}
\subsection{Model filtering}
\label{sec:ModRP_discuss}
We highlight the necessity for proper model filtering when computing the forward model. This ensures that the forward model accurately represents the noiseless data, and thus allows for the accurate retrieval of injected model parameters. This model filtering is also necessary for the accurate retrieval of real planetary signals. For the injected data preprocessed with \textsc{SysRem}, we performed a full \textsc{SysRem}/PCA filtering to the forward model \citep{Gibson_2022} each time we computed the log-likelihood for a given set of model parameters within the MCMC. Similarly, the division by the median out-of-transit spectrum for the injected data preprocessed with \textsc{molecfit} also distorted the injected signal, and must be accounted for. The division of the planetary signal by stellar lines caused a distortion in the line shape/position, which materialised as an anomalous offset from the injected $K_{\rm P}$ value (see Fig.~\ref{Maps}). Therefore, without accounting for this step in the computation of the forward model, we risked an inaccurate retrieval of the injected parameters.
\subsection{Atmospheric retrieval comparison}
\label{sec:AtmoRetDiscuss}
\subsubsection{P = 2 days}
\label{sec:AtmoRetDiscuss_P2}
For DS1, both reduction methods resulted in accurate recovered values for the atmospheric species' abundances relative to Fe (see Fig.~\ref{Obs1_P2_relabuns}). However, both methods also caused the recovery of a slightly higher Na/Fe value, despite still having overlapping probability mass with the injected value. The \textsc{molecfit} preprocessing method also caused a decrease in the recovered OH/Fe abundance with respect to the injected value, caused by a lower recovered OH abundance. OH appears in the Earth's atmosphere as emission, where \textsc{molecfit} does not correct for atmospheric emission, which explains any biases introduced by its non-removal. This issue would potentially be resolved with the use of the sky-subtracted products provided by the DRS. Furthermore, the H$_2$O/Fe abundance is broader for the \textsc{molecfit} case, which could be due to \textsc{SysRem}'s ability to model and remove additional systematics present in the data, which also effect the exoplanetary signal, and thus ultimately provide better constraints on the injected abundances. Both methods also recovered the injected vertical $T$-$P$ profile accurately (see Figs.~\ref{Obs1_P2_Sys_corner} \&~\ref{Obs1_P2_Mol_corner}). It is worth noting that the forward model is truncated below the cloud deck pressure ($P_{\rm cloud}$), thus the atmospheric retrieval is unable to constrain an accurate $T$-$P$ profile below this value (see Fig.~\ref{Obs1_P2_Sys_corner} for reference). Furthermore, the choice of a parameterised $T$-$P$ profile will bias the recovery, as the injected and recovered parameterisation will be identical. An alternative case which is not investigated here is a self-consistent $T$-$P$ profile injected into our datasets, which is then recovered with a parameterised profile, to highlight any potential biases in the chosen parameterisation.

For DS2, we again saw an accurate recovery of the injected model parameters for both data reduction methods (see Fig.~\ref{Obs2_P2_relabuns}). The \textsc{SysRem} preprocessing method did, however, cause a slightly higher OH/Fe value to be recovered, due to an underestimation of the injected Fe abundance, though still consistent with the injected value with 2.5$\sigma$ (see Fig.~\ref{Obs2_P2_Sys_corner}). Both methods again recovered the injected vertical $T$-$P$ profile accurately (see Figs.~\ref{Obs2_P2_Sys_corner} \&~\ref{Obs2_P2_Mol_corner}).

Finally, for DS3, the injected model parameters were accurately recovered, with each reduction method performing similarly. There was, however, broad constraints on the H$_2$O/Fe ratio for the \textsc{molecfit} preprocessing method, with a local maxima in the posterior distribution consistent with the injected value. In contrast, the \textsc{SysRem} preprocessing method recovers a well-constrained H$_2$O/Fe posterior at the injected value (see Fig.~\ref{Obs3_P2_relabuns}). Both reduction methods also caused the injected $T$-$P$ profiles to be recovered within 2$\sigma$ (see Figs.~\ref{Obs3_P2_Sys_corner} \&~\ref{Obs3_P2_Mol_corner}), with the \textsc{SysRem} preprocessing resulting in more precise constraints on the $T$-$P$ profile overall.
\begin{figure}
    \includegraphics[width=\columnwidth]{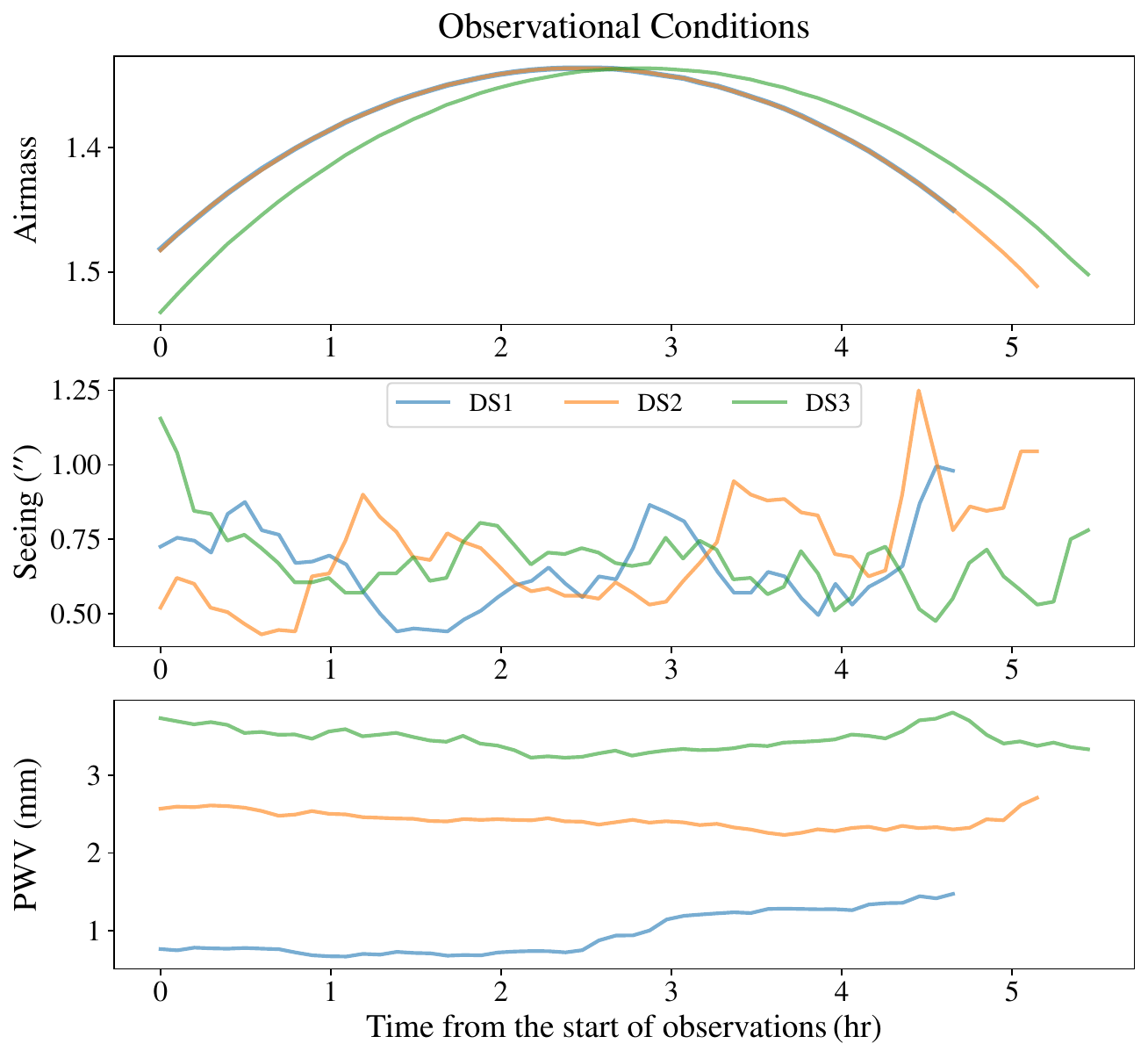}
    \caption{Various observational conditions of each of the datasets. The top panel shows the airmass variation throughout the observations. The middle panel shows the seeing variation throughout the observations. The bottom panel shows the precipitable water vapour (PWV) variation throughout the observations.} 
    \label{ObsConds}
\end{figure}
\subsubsection{P = 12 days}
For DS1, the \textsc{molecfit} preprocessing method resulted in an accurate recovery of the abundances, as well as the $T$-$P$ profile (see Figs.~\ref{Obs1_P12_relabuns} \& \ref{Obs1_P12_Mol_corner}). The recovered H$_2$O abundance shows broad constraints, with a local maxima at the injected value. However, for the \textsc{SysRem} preprocessing case, we found an unconstrained H$_2$O/Fe value, caused by an unconstrained posterior for the H$_2$O abundance. This is likely due to the fact that the planetary H$_2$O lines exist at locations close to the telluric H$_2$O lines, albeit shifted by $\Delta v_{\rm sys}$ and drifting in time due to $\boldsymbol{v}_{\rm \textbf{p}}$ (see Eq.~\ref{eq:planet_vel}). Therefore, as a PCA-like algorithm such as \textsc{SysRem} fits these regions with many telluric lines, it inadvertently fits/removes the planetary lines as well. This is particularly true in this case of P12 ($|K_{\rm p}|$ = 57.29 km s$^{-1}$), as the slow-moving planetary signal's wavelength shift during a transit is less prominent (i.e., it is more stationary with respect to time), causing it to be `targeted' more effectively by these fitting algorithms. It is also worth noting that the optimum number of \textsc{SysRem} iterations (10) was greater for this injection and recovery test when compared to DS2 and DS3, despite the fact that DS1 benefited from the lowest amount of precipitable water vapour (PWV) of all three nights of observations (see Fig.~\ref{ObsConds}). This also caused the recovered $T$-$P$ profile to differ from the injected value by greater than 3$\sigma$ in this case (see Fig.~\ref{Obs1_P12_Sys_corner}). This injection scenario and dataset was also tested with a fewer number of \textsc{SysRem} iterations (5) in order to mitigate this issue, however the unconstrained H$_2$O/Fe posterior remained.

For DS2, both reduction methods performed similarly, likely due to the relatively low number of \textsc{SysRem} iterations used. For both, we found V/Fe, OH/Fe, and H$_2$O/Fe values recovered slightly higher than, but still in agreement with, the injected value, due to an Fe abundance recovered below the injected value (see Fig.~\ref{Obs2_P12_relabuns}). Despite this, both methods recovered the injected $T$-$P$ profile accurately (see Figs.~\ref{Obs2_P12_Sys_corner} \&~\ref{Obs2_P12_Mol_corner}).

Similarly to the P2 forward model injected into DS3, we found a broad heavy-tailed H$_2$O/Fe distribution for the \textsc{molecfit} preprocessing case, with a prominent local maxima at the injected value (see Fig.~\ref{Obs3_P12_relabuns}). The \textsc{SysRem} preprocessing method, however, resulted in a more precise H$_2$O/Fe value at the injected value, albeit less accurate than the \textsc{molecfit} value. The more precise abundance constraints provided by \textsc{SysRem} with respect to \textsc{molecfit} in this dataset, for both injection scenarios, hinted at potential additional systematics present in the data which were affecting the injected exoplanet signal, and which were then modelled and removed more effectively by \textsc{SysRem}. Furthermore, the PWV measured for this dataset was the highest of all three nights of observations. The remaining injected abundances were also recovered accurately by both reduction methods. Both also recover the injected $T$-$P$ profile accurately, with the \textsc{molecfit} method providing slightly more stringent constraints (see Figs.~\ref{Obs3_P12_Sys_corner} \&~\ref{Obs3_P12_Mol_corner}).

\section{Conclusions}
\label{sec:Conclusions}
In this work we have presented numerous injection and recovery tests of exoplanetary signals in transmission, on two distinct orbital periods of P = 2 and P = 12 days. The purpose of these distinct orbital architectures was to test the limitations of the typical high-resolution reduction methods used on planetary signals with very different orbital velocities and thus different Doppler shift of their respective signals in-transit. Primarily, we wished to assess and compare the efficacy of the two most common telluric removal methods used in the literature--PCA-like algorithms such as \textsc{SysRem} and the telluric modelling software \textsc{molecfit}, and their effects on the retrieval of atmospheric parameters. 

We found that for a Jupiter-like planet injected at P = 2 days, the two telluric removal methods performed similarly, providing accurate recovery of the exoplanetary signal with precise constraints. However, we found that \textsc{SysRem} typically performed better for species that are also present in the Earth’s atmosphere. For the H$_2$O/Fe abundance, we find a slightly more precise constraint when the data was preprocessed with \textsc{SysRem} across all three datasets. This is notably true for DS3, where the heavy-tailed H$_2$O abundance seen for the \textsc{molecfit} reduction causes broad, heavy-tailed constraints on H$_2$O/Fe.

Furthermore, for a Jupiter-like planet injected at P = 12 days, we saw a well constrained H$_2$O/Fe value for DS1 in the \textsc{molecfit} case, whereas for \textsc{SysRem}, the H$_2$O signal was effectively removed, causing an unconstrained H$_2$O/Fe posterior. This behaviour is expected, as a slower moving planetary signal will be targeted and distorted more effectively by PCA-like algorithms, as they appear more stationary in time relative to the slower moving, close-in planetary signals. However, this behaviour was not ubiquitous across all three datasets we investigated, and in the case of DS3, we saw a more precise but less accurate H$_2$O/Fe constraint when we used \textsc{SysRem} to treat the tellurics. This result is consistent with the findings of \cite{Cheverall_2024} where they found it practicable to recover injected signals of the sub-Neptune TOI-732c from IGRINS data when using a PCA-based telluric removal method, despite the planet's low radial velocity shift during transit. Both methods performed similarly for DS2, with accurate and precise retrievals of the injected signals. 

It is with the above results and conclusions in mind that we would state that there is no definitive, ``one size fits all" method for modelling/removing telluric contamination from high-resolution optical datasets. The delicate and often aggressive nature of high-resolution data filtering means that the utmost care must be taken in ensuring real exoplanetary signals are not only preserved, but any inevitable distortions caused by these reduction methods must be minimised and accounted for when computing the forward model. This is evident, we found, even for cross-correlation analysis (see Sect.~\ref{sec:ModRP_discuss}). Ideally, the impact, or lack thereof, of high-resolution reduction frameworks on real exoplanetary data would be highlighted via injection and recovery tests on a case-by-case basis. Furthermore, a technique which was not explored as part of this study, but would potentially be of interest in the future, is the combination of multiple telluric modelling/removal methods. This would be straightforward to implement in the framework outlined in this work, with radiative transfer modelling such as \textsc{molecfit} handling dominant telluric contamination, and PCA-like algorithms such as \textsc{SysRem} handling additional time-invariant systematics. 

Although this study presented analysis of an idealised test case, free from additional in-transit contamination such as the RM and CLV effects, it nonetheless provides a test bed from which to extrapolate the different biases introduced by the most common telluric removal methods on various atomic and molecular absorption signals in the optical wavelength regime. Going forward, a similar study in the near-infrared/infrared with instruments on the VLT such as CRIRES+ would also be of great interest. At these wavelengths, telluric contamination - particularly from H$_2$O - dominates and can often saturate the absorption spectrum. However, modelling and removing these telluric bands is of the utmost importance in the search for telluric signatures in exoplanetary atmospheres from the ground. 

\begin{acknowledgements}
This work relied on observations collected at the European Organisation for Astronomical Research in the Southern Hemisphere under ESO programme 106.21BV.005. The author(s) gratefully acknowledge support from Science Foundation Ireland and the Royal Society in the form of a Research Fellows Enhancement Award. CM and ES would like to acknowledge financial support from ESO's Scientific Support Discretionary Fund (SSDF) that enabled the realisation of this study. We are grateful to the developers of the NumPy, SciPy, Matplotlib, iPython, corner, petitRADTRANS, and Astropy packages, which were used extensively in this work \citep{Numpy,Scipy,matplotlib,ipython,corner,Molliere_2019}.
\end{acknowledgements}

%
\bibliographystyle{aa} 
\bibliography{Bibliography.bib} 

\begin{appendix}
\onecolumn
\section{Injected transmission spectra}
\label{appendix_a}
\begin{figure*}[!htbp]
    \centering
    \includegraphics[width=\textwidth]{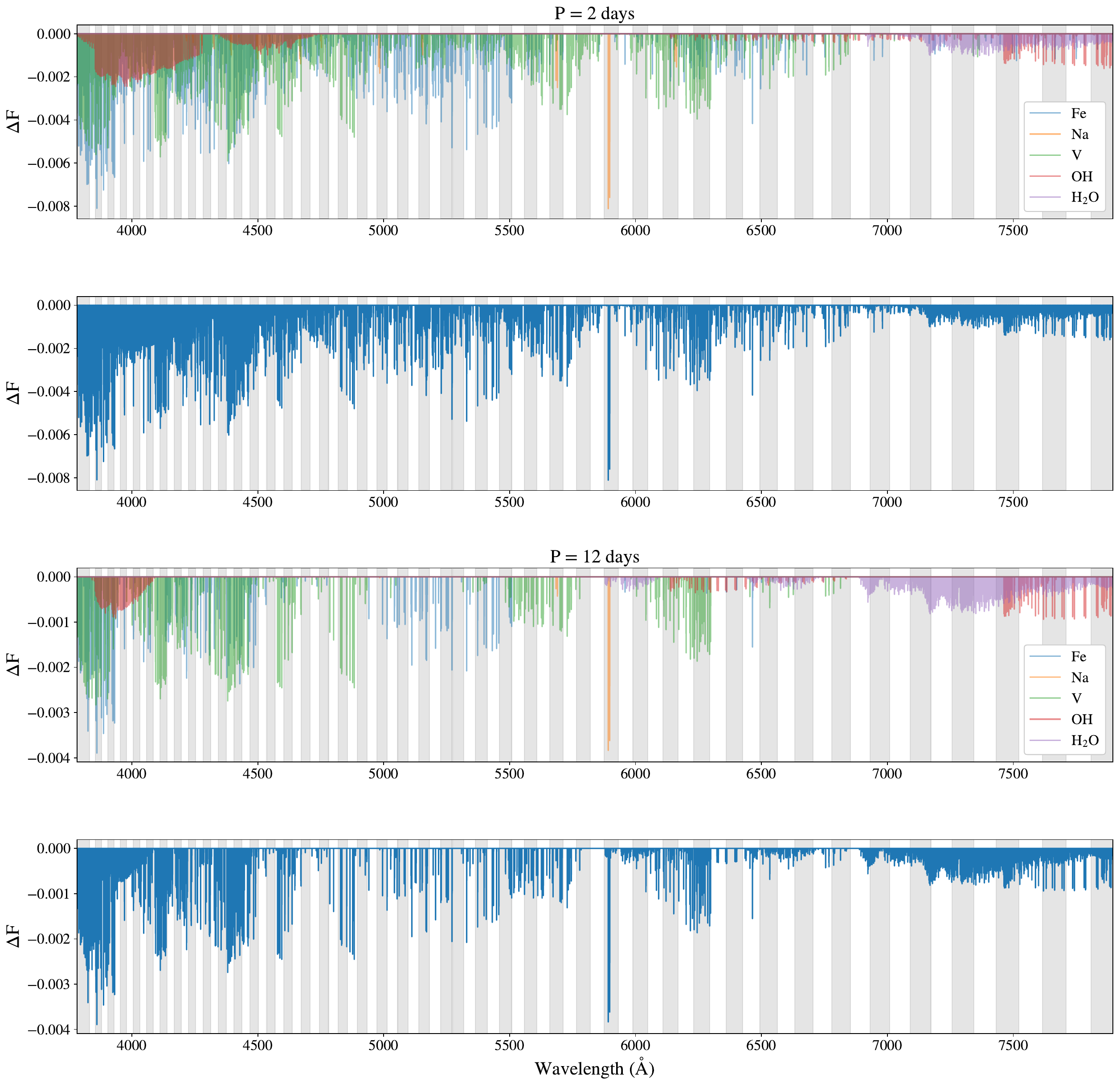}
        \caption{Model transmission spectra injected into each of the datasets. The top two panels show both the individual species' contributions, as well as the total unbroadened spectrum for P = 2 days. The bottom two panels show both the individual species' contributions, as well as the total unbroadened spectrum for P = 12 days. The alternating white and grey shaded regions in each of the panels are the individual ESPRESSO orders after pairing, described in Sect.~\ref{sec:molecfit}}.
        \label{Injected_models}
\end{figure*}
\newpage
\section{Additional relative abundance comparisons}
\label{appendix_b}
\begin{figure*}[!htbp]
    \centering
    \includegraphics[width=0.75\textwidth]{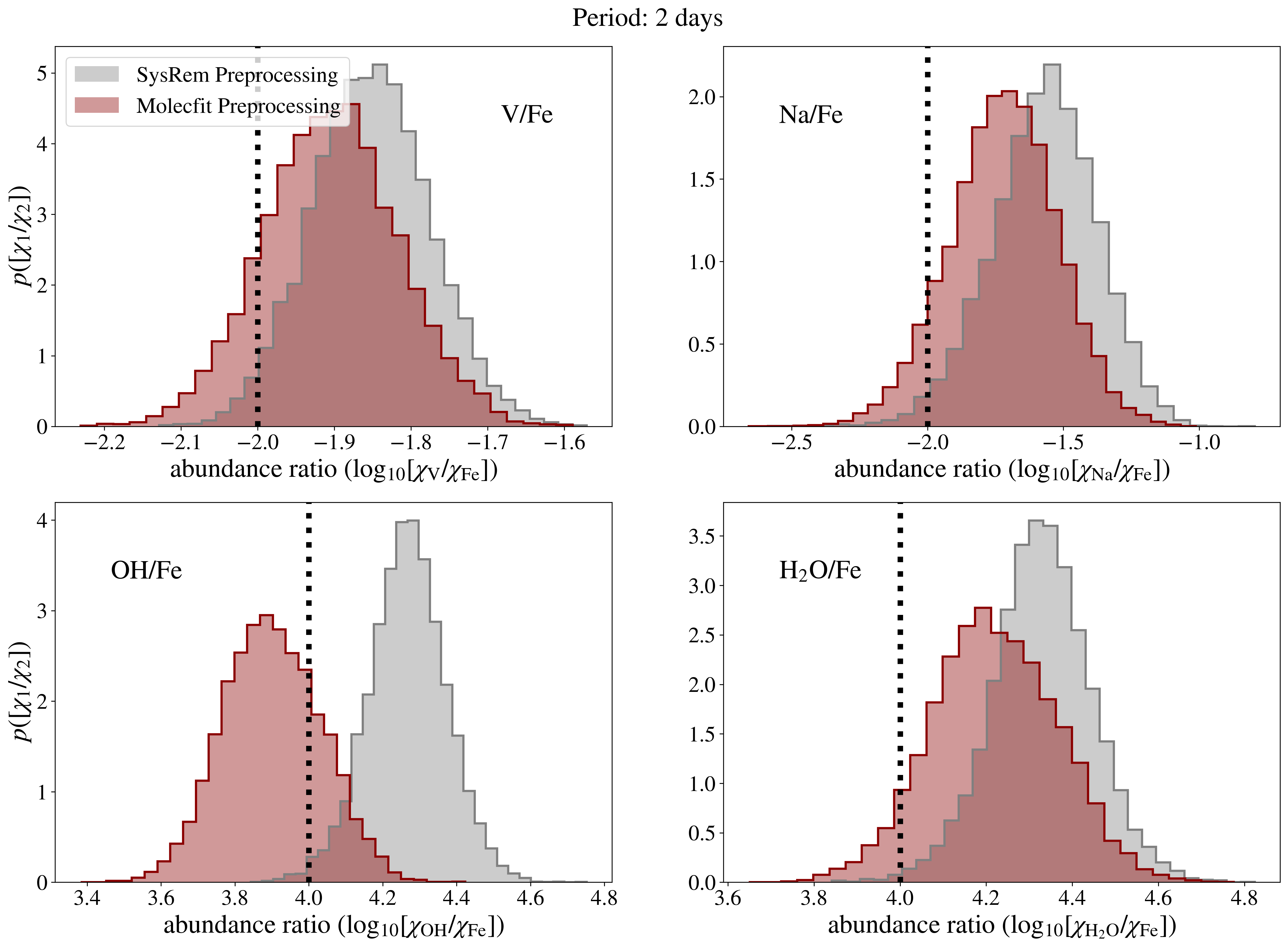}
        \caption{Similar to Fig.~\ref{Obs1_P2_relabuns}, for DS2.} 
        \label{Obs2_P2_relabuns}
\end{figure*}
\begin{figure*}[!htbp]
    \centering
    \includegraphics[width=0.75\textwidth]{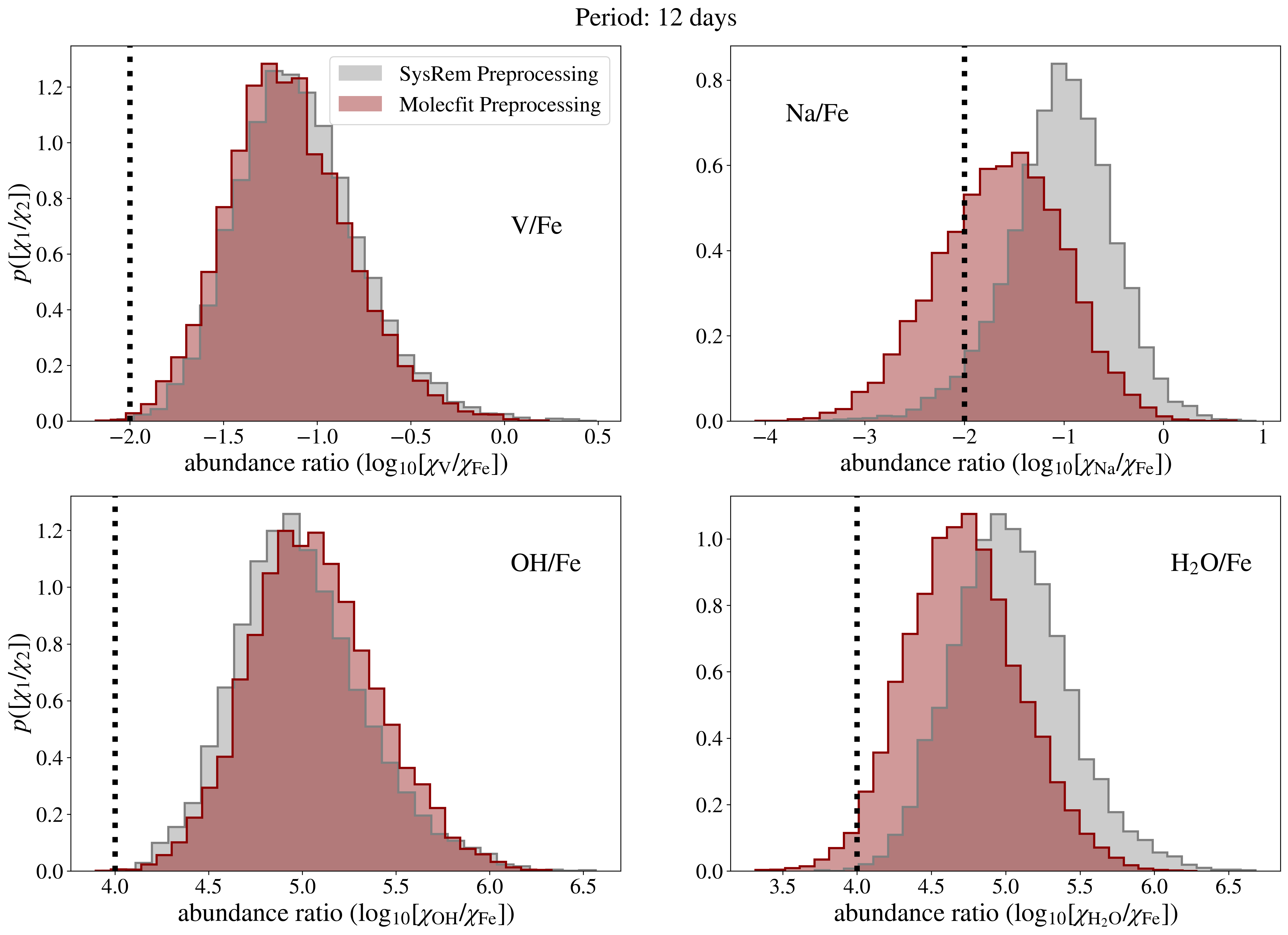}
        \caption{Similar to Fig.~\ref{Obs1_P12_relabuns}, for DS2.} 
        \label{Obs2_P12_relabuns}
\end{figure*}
\begin{figure*}[!htbp]
    \centering
    \includegraphics[width=0.75\textwidth]{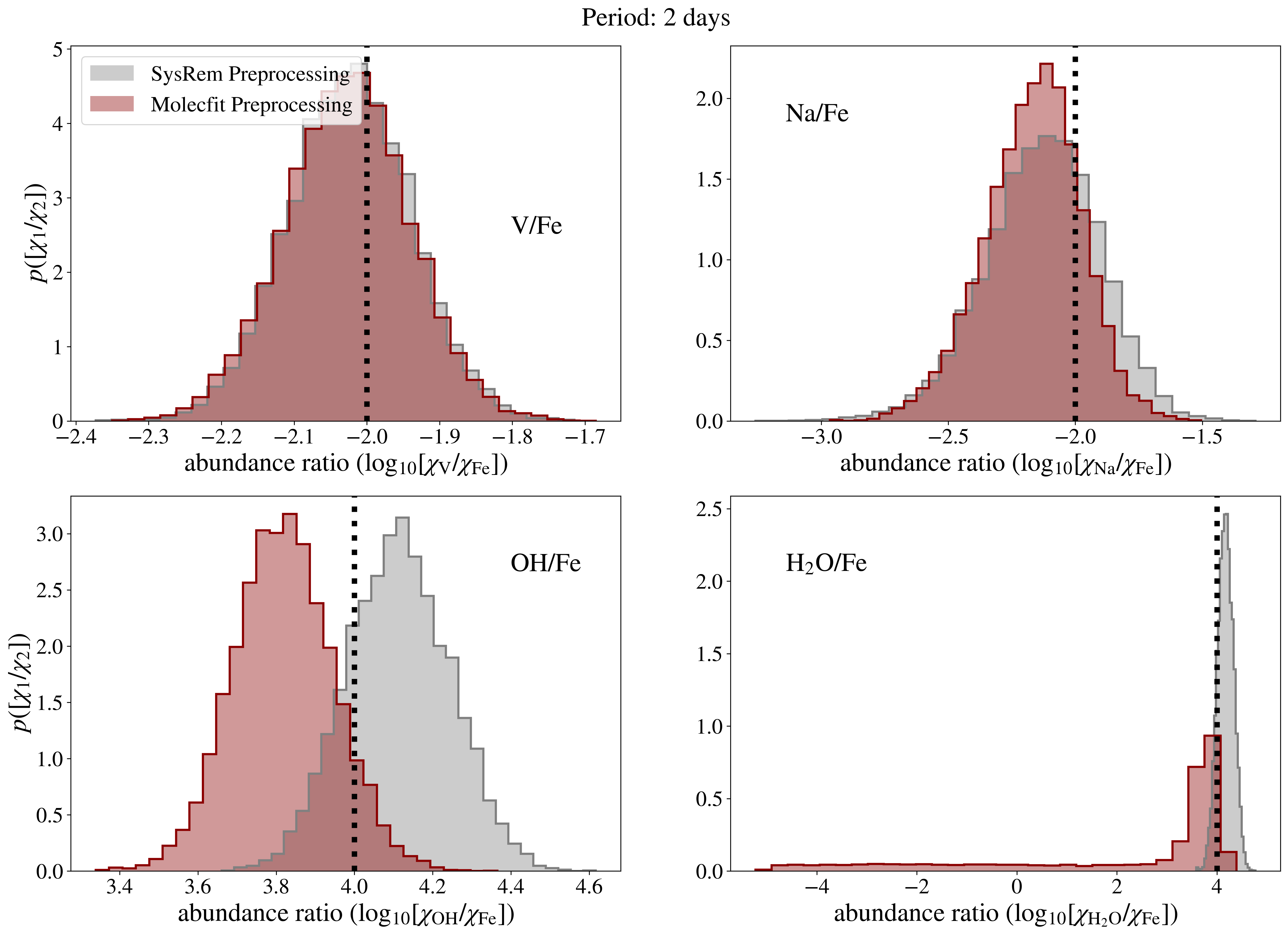}
        \caption{Similar to Fig.~\ref{Obs1_P2_relabuns}, for DS3.} 
        \label{Obs3_P2_relabuns}
\end{figure*}
\begin{figure*}[!htbp]
    \centering
    \includegraphics[width=0.75\textwidth]{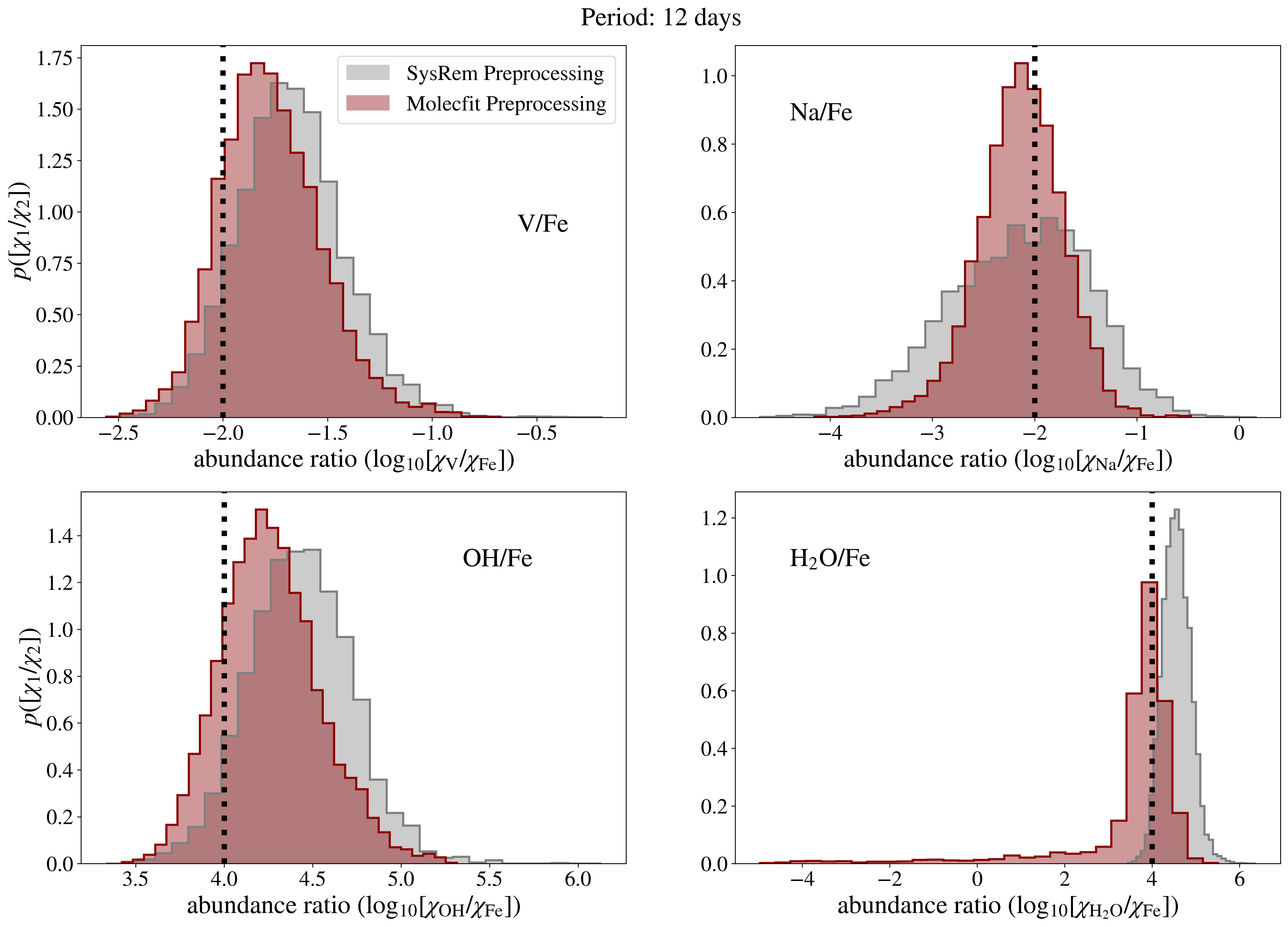}
        \caption{Similar to Fig.~\ref{Obs1_P12_relabuns}, for DS3.} 
        \label{Obs3_P12_relabuns}
\end{figure*}
\clearpage
\section{Atmospheric retrieval corner plots}
\label{appendix_c}
\begin{figure*}[!htbp]
    \centering
    \includegraphics[width=\textwidth]{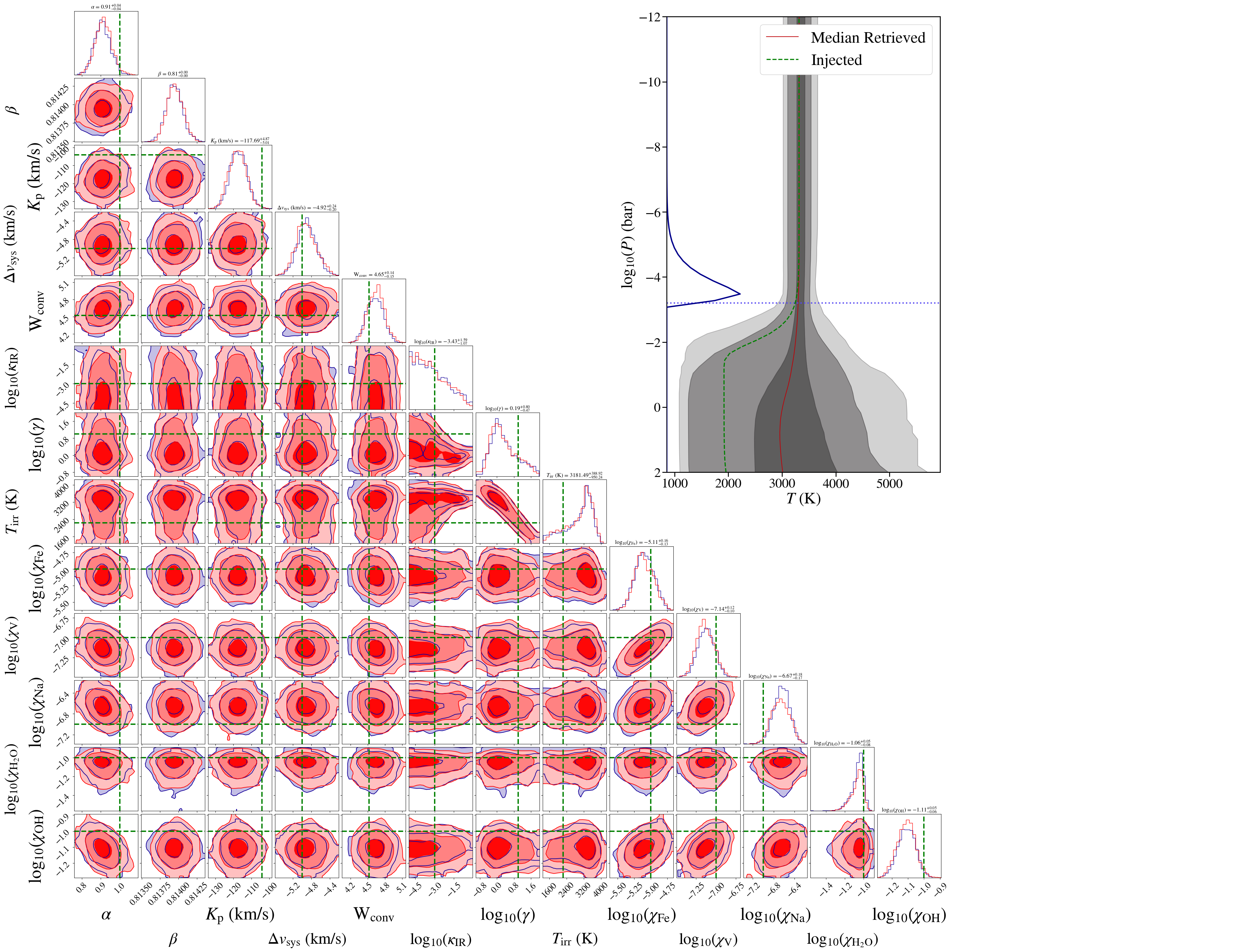}
        \caption{Atmospheric retrieval results for P2 injected into DS1 and preprocessed with \textsc{SysRem}, with the 1D and 2D marginalised posterior distributions of each of the model parameters displayed. The red and blue posterior distributions represent independent subchains, of the same MCMC chain, both converging to similar distributions. The injected values are shown as green vertical/horizontal dashed lines. The $T$-$P$ profile shown on the right was computed from 10,000 random samples of the MCMC, where the solid red curve shows the median profile, and the grey shaded regions show the 1$\sigma$, 2$\sigma$, and 3$\sigma$ contours. The injected profile is shown as a green dashed curve. The cloud deck, located at $\log_{10}(P_{\rm cloud})$ is highlighted as a blue dotted line. Below this value, the model is truncated, and thus the model inference breaks down. The contribution curve is also shown in blue.} 
        \label{Obs1_P2_Sys_corner}
\end{figure*}
\begin{figure*}[!htbp]
    \centering
    \includegraphics[width=\textwidth]{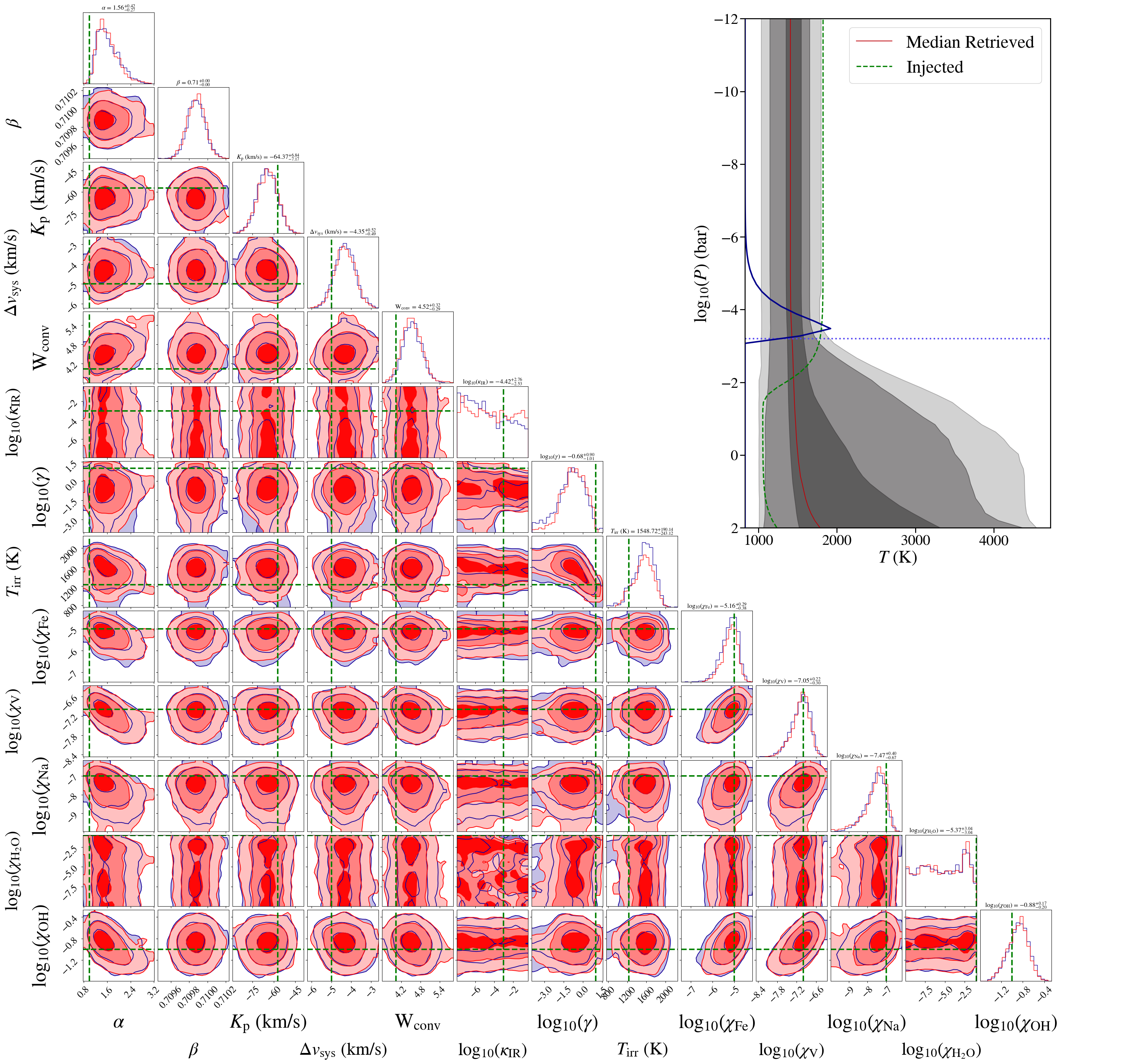}
        \caption{Similar to Fig.~\ref{Obs1_P2_Sys_corner}, for P12.} 
        \label{Obs1_P12_Sys_corner}
\end{figure*}
\begin{figure*}[!htbp]
    \centering
    \includegraphics[width=\textwidth]{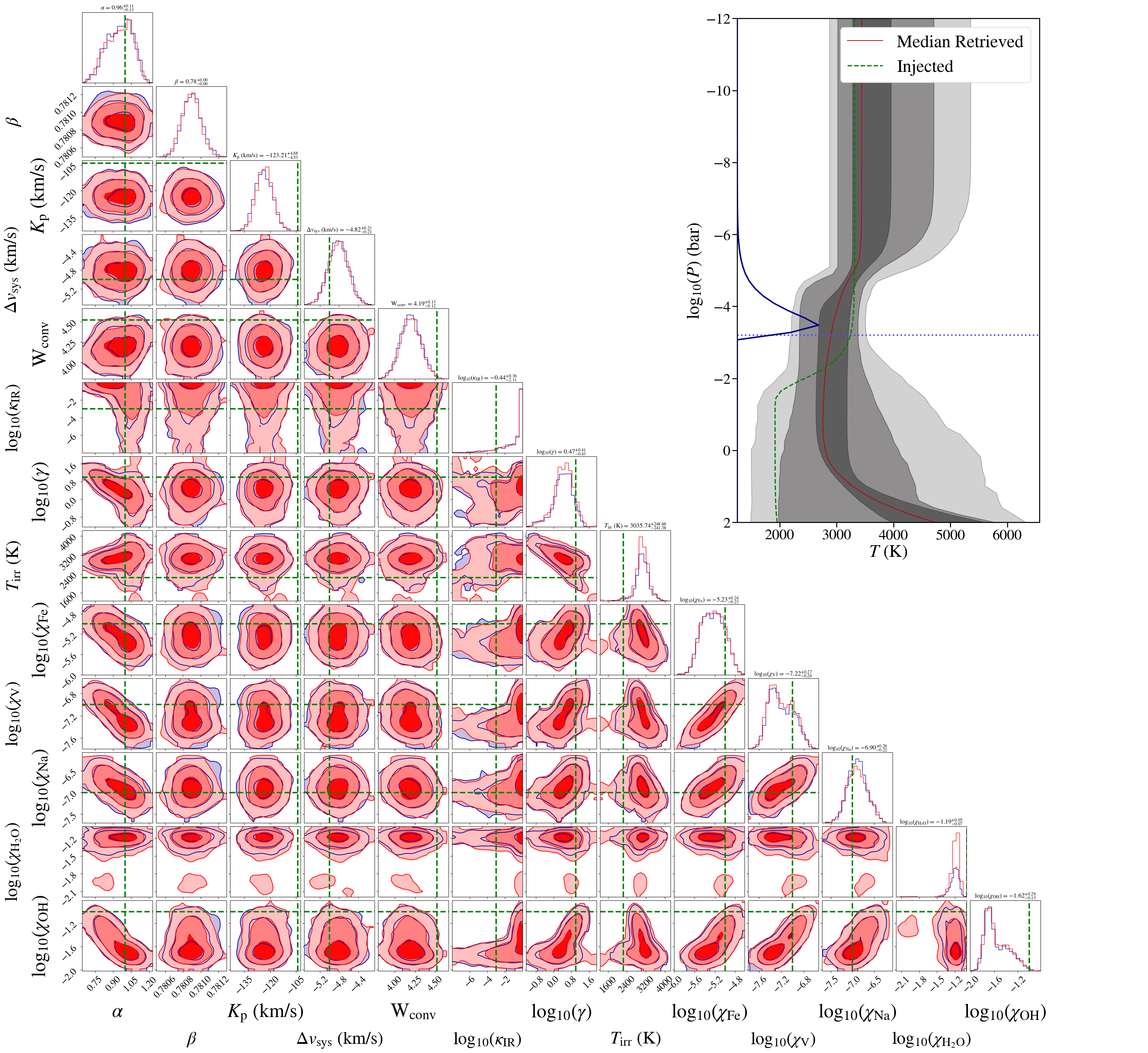}
        \caption{Atmospheric retrieval results for P2 injected into DS1 and preprocessed with \textsc{molecfit}, with the 1D and 2D marginalised posterior distributions of each of the model parameters displayed. The red and blue posterior distributions represent independent subchains, of the same MCMC chain, both converging to similar distributions. The injected values are shown as green vertical/horizontal dashed lines. The $T$-$P$ profile shown on the right was computed from 10,000 random samples of the MCMC, where the solid red curve shows the median profile, and the grey shaded regions show the 1$\sigma$, 2$\sigma$, and 3$\sigma$ contours. The injected profile is shown as a green dashed curve. The cloud deck, located at $\log_{10}(P_{\rm cloud})$ is highlighted as a blue dotted line. Below this value, the model is truncated, and thus the model inference breaks down. The contribution curve is also shown in blue.} 
        \label{Obs1_P2_Mol_corner}
\end{figure*}
\begin{figure*}[!htbp]
    \centering
    \includegraphics[width=\textwidth]{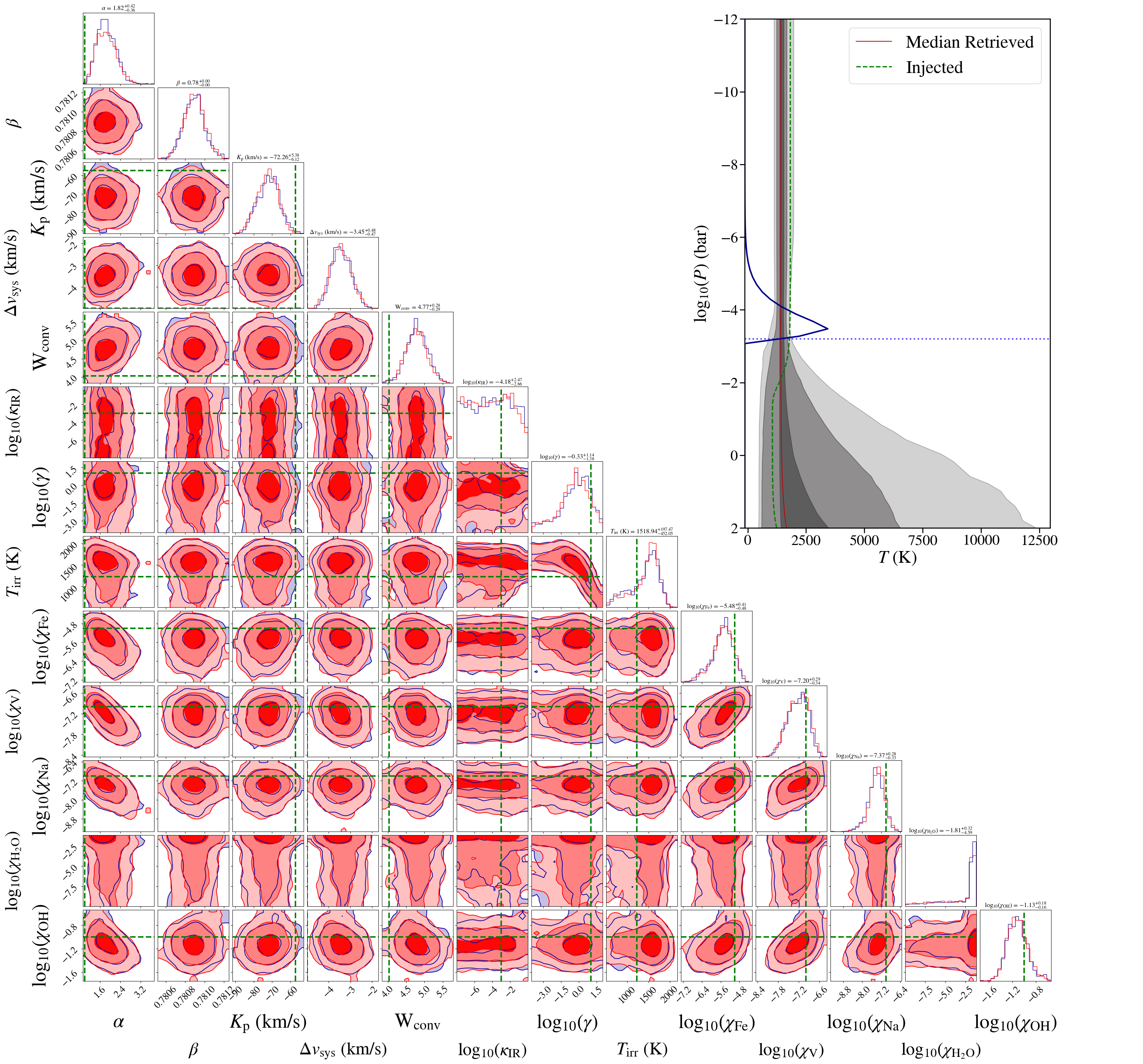}
        \caption{Similar to Fig.~\ref{Obs1_P2_Mol_corner}, for P12.} 
        \label{Obs1_P12_Mol_corner}
\end{figure*}

\begin{figure*}[!htbp]
    \centering
    \includegraphics[width=\textwidth]{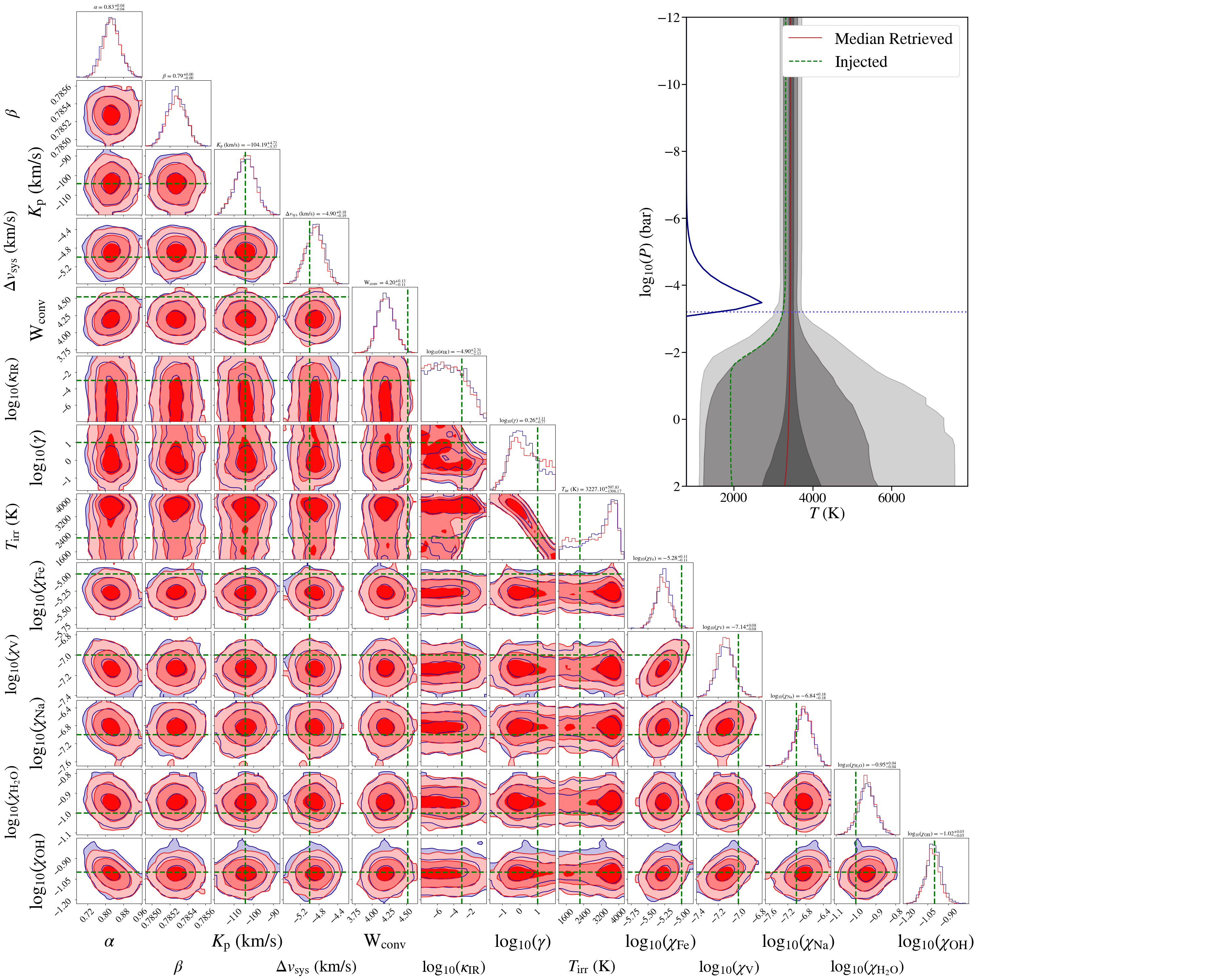}
        \caption{Atmospheric retrieval results for P2 injected into DS2 and preprocessed with \textsc{SysRem}, with the 1D and 2D marginalised posterior distributions of each of the model parameters displayed. The red and blue posterior distributions represent independent subchains, of the same MCMC chain, both converging to similar distributions. The injected values are shown as green vertical/horizontal dashed lines. The $T$-$P$ profile shown on the right was computed from 10,000 random samples of the MCMC, where the solid red curve shows the median profile, and the grey shaded regions show the 1$\sigma$, 2$\sigma$, and 3$\sigma$ contours. The injected profile is shown as a green dashed curve. The cloud deck, located at $\log_{10}(P_{\rm cloud})$ is highlighted as a blue dotted line. Below this value, the model is truncated, and thus the model inference breaks down. The contribution curve is also shown in blue.} 
        \label{Obs2_P2_Sys_corner}
\end{figure*}
\begin{figure*}[!htbp]
    \centering
    \includegraphics[width=\textwidth]{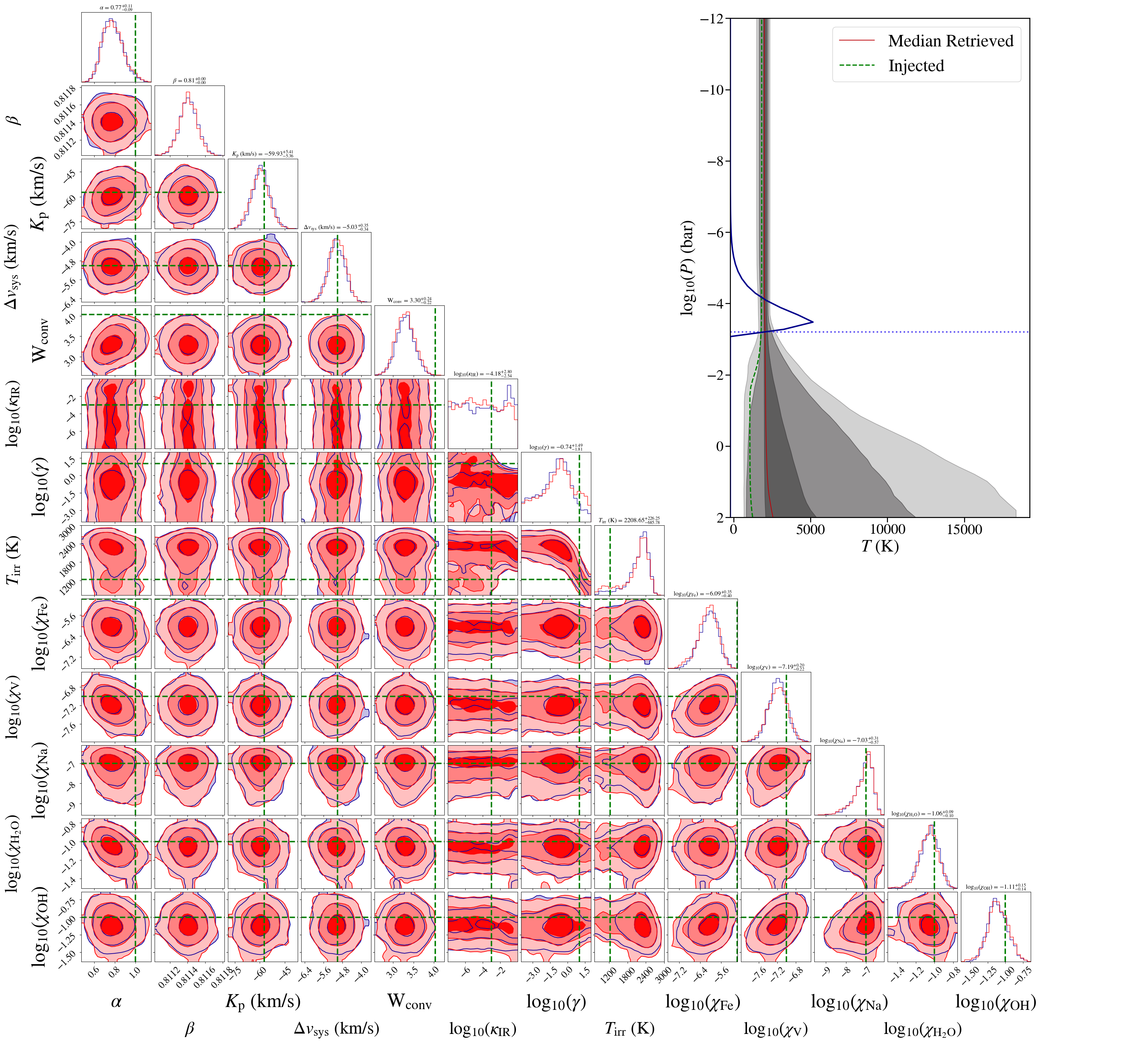}
        \caption{Similar to Fig.~\ref{Obs2_P2_Sys_corner}, for P12.} 
        \label{Obs2_P12_Sys_corner}
\end{figure*}
\begin{figure*}[!htbp]
    \centering
    \includegraphics[width=\textwidth]{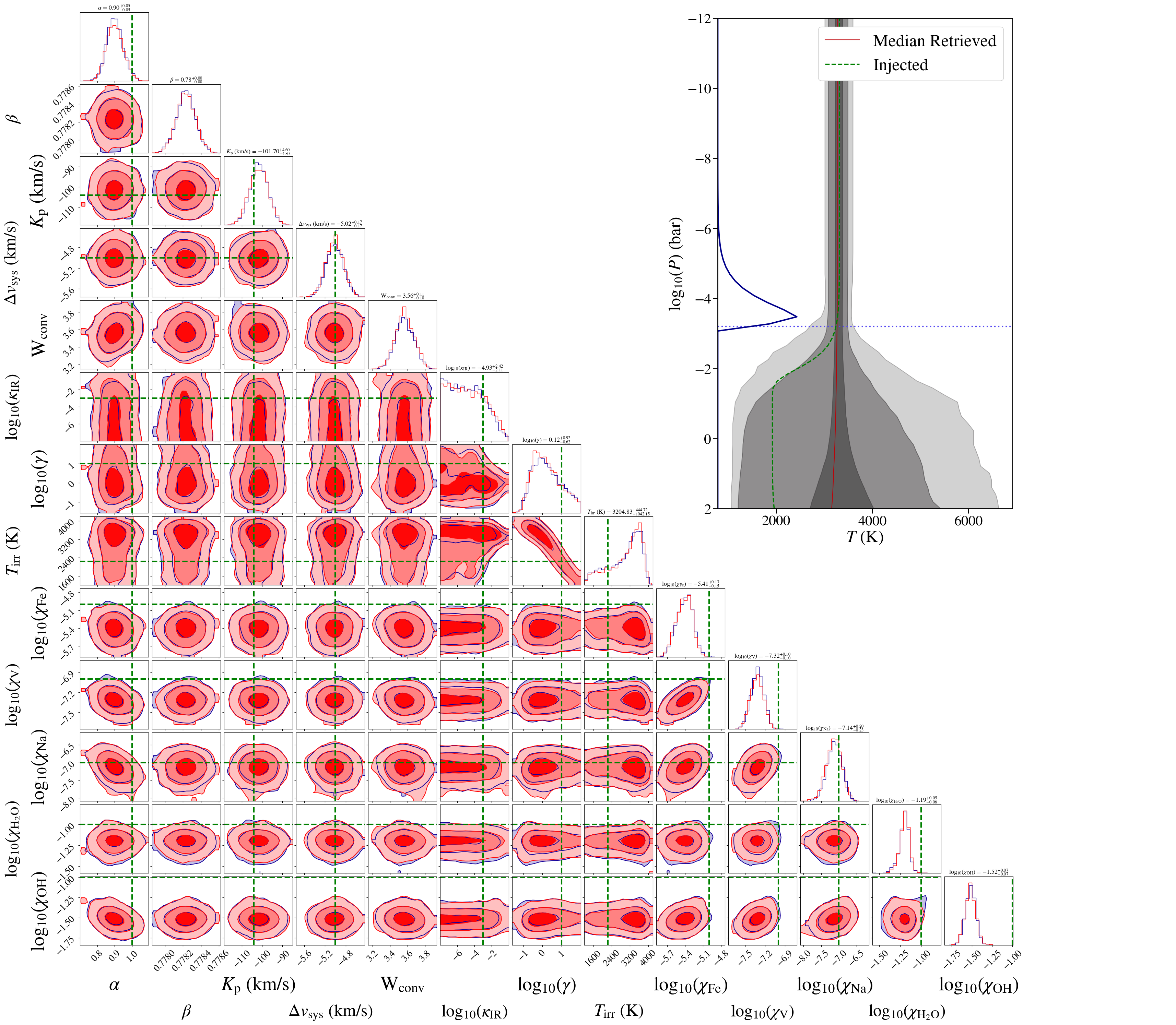}
        \caption{Atmospheric retrieval results for P2 injected into DS2 and preprocessed with \textsc{molecfit}, with the 1D and 2D marginalised posterior distributions of each of the model parameters displayed. The red and blue posterior distributions represent independent subchains, of the same MCMC chain, both converging to similar distributions. The injected values are shown as green vertical/horizontal dashed lines. The $T$-$P$ profile shown on the right was computed from 10,000 random samples of the MCMC, where the solid red curve shows the median profile, and the grey shaded regions show the 1$\sigma$, 2$\sigma$, and 3$\sigma$ contours. The injected profile is shown as a green dashed curve. The cloud deck, located at $\log_{10}(P_{\rm cloud})$ is highlighted as a blue dotted line. Below this value, the model is truncated, and thus the model inference breaks down. The contribution curve is also shown in blue.} 
        \label{Obs2_P2_Mol_corner}
\end{figure*}
\begin{figure*}[!htbp]
    \centering
    \includegraphics[width=\textwidth]{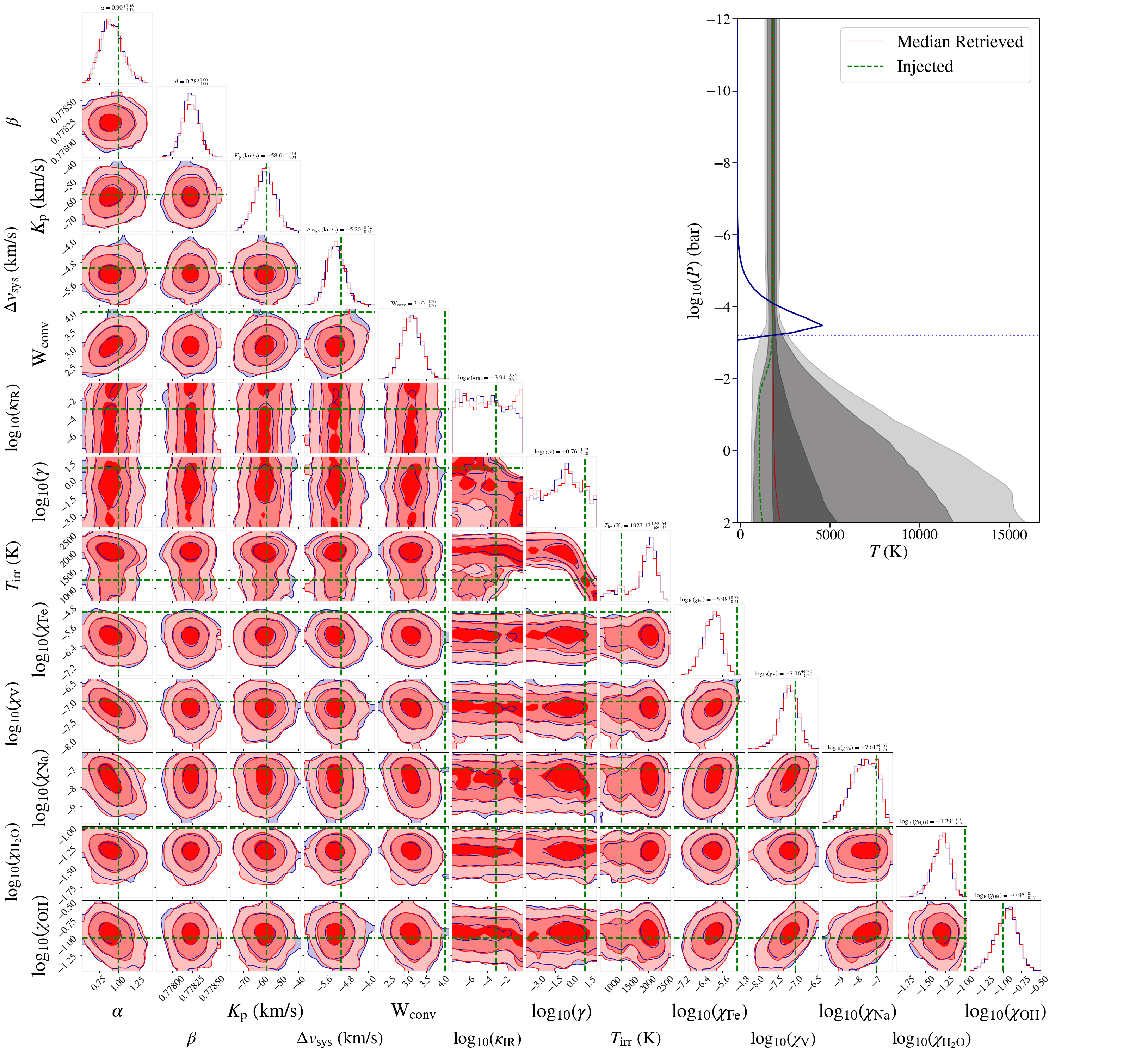}
        \caption{Similar to Fig.~\ref{Obs2_P2_Mol_corner}, for P12.} 
        \label{Obs2_P12_Mol_corner}
\end{figure*}

\begin{figure*}[!htbp]
    \centering
    \includegraphics[width=\textwidth]{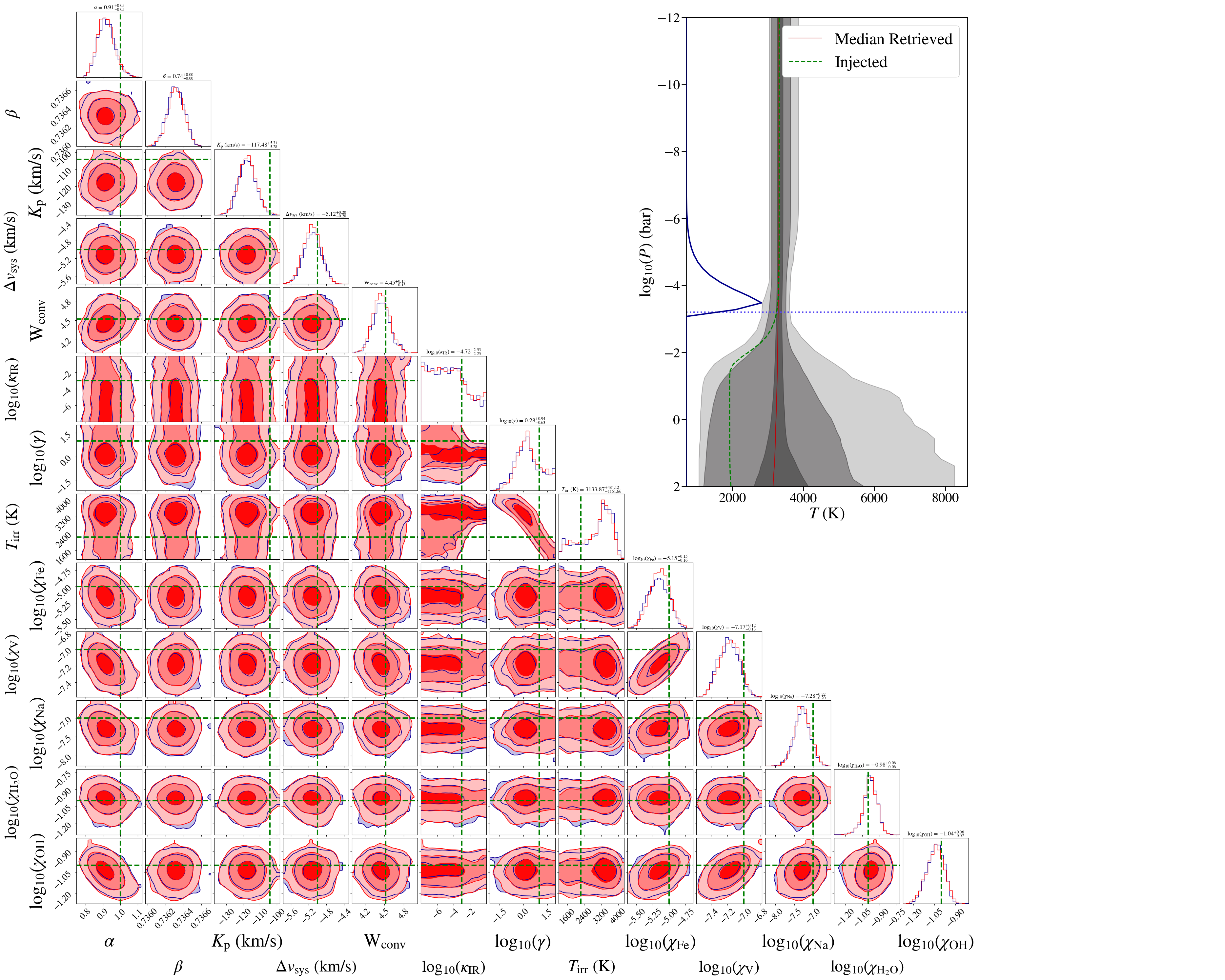}
        \caption{Atmospheric retrieval results for P2 injected into DS3 and preprocessed with \textsc{SysRem}, with the 1D and 2D marginalised posterior distributions of each of the model parameters displayed. The red and blue posterior distributions represent independent subchains, of the same MCMC chain, both converging to similar distributions. The injected values are shown as green vertical/horizontal dashed lines. The $T$-$P$ profile shown on the right was computed from 10,000 random samples of the MCMC, where the solid red curve shows the median profile, and the grey shaded regions show the 1$\sigma$, 2$\sigma$, and 3$\sigma$ contours. The injected profile is shown as a green dashed curve. The cloud deck, located at $\log_{10}(P_{\rm cloud})$ is highlighted as a blue dotted line. Below this value, the model is truncated, and thus the model inference breaks down. The contribution curve is also shown in blue.} 
        \label{Obs3_P2_Sys_corner}
\end{figure*}
\begin{figure*}[!htbp]
    \centering
    \includegraphics[width=\textwidth]{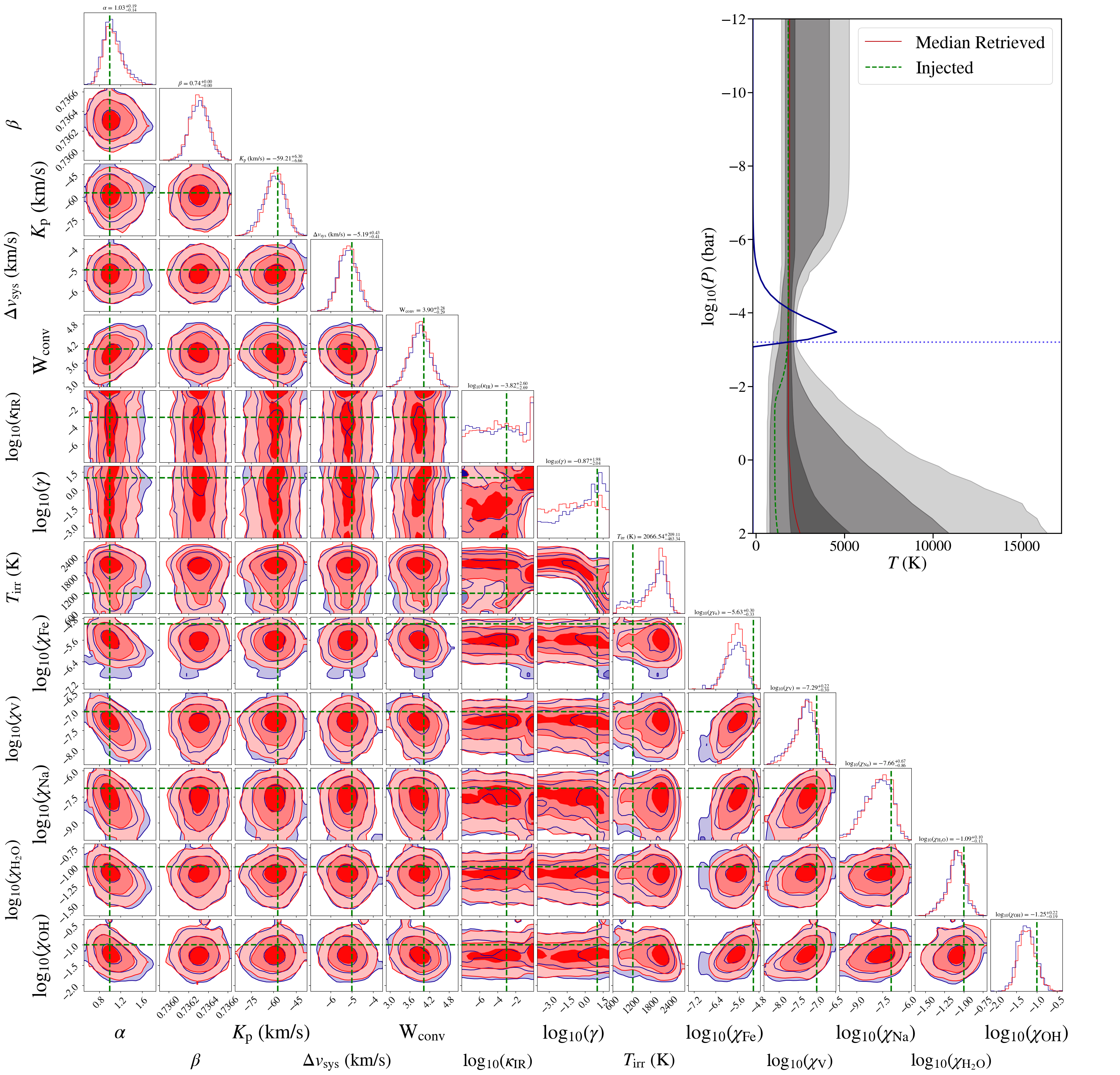}
        \caption{Similar to Fig.~\ref{Obs3_P2_Sys_corner}, for P12.} 
        \label{Obs3_P12_Sys_corner}
\end{figure*}
\begin{figure*}[!htbp]
    \centering
    \includegraphics[width=\textwidth]{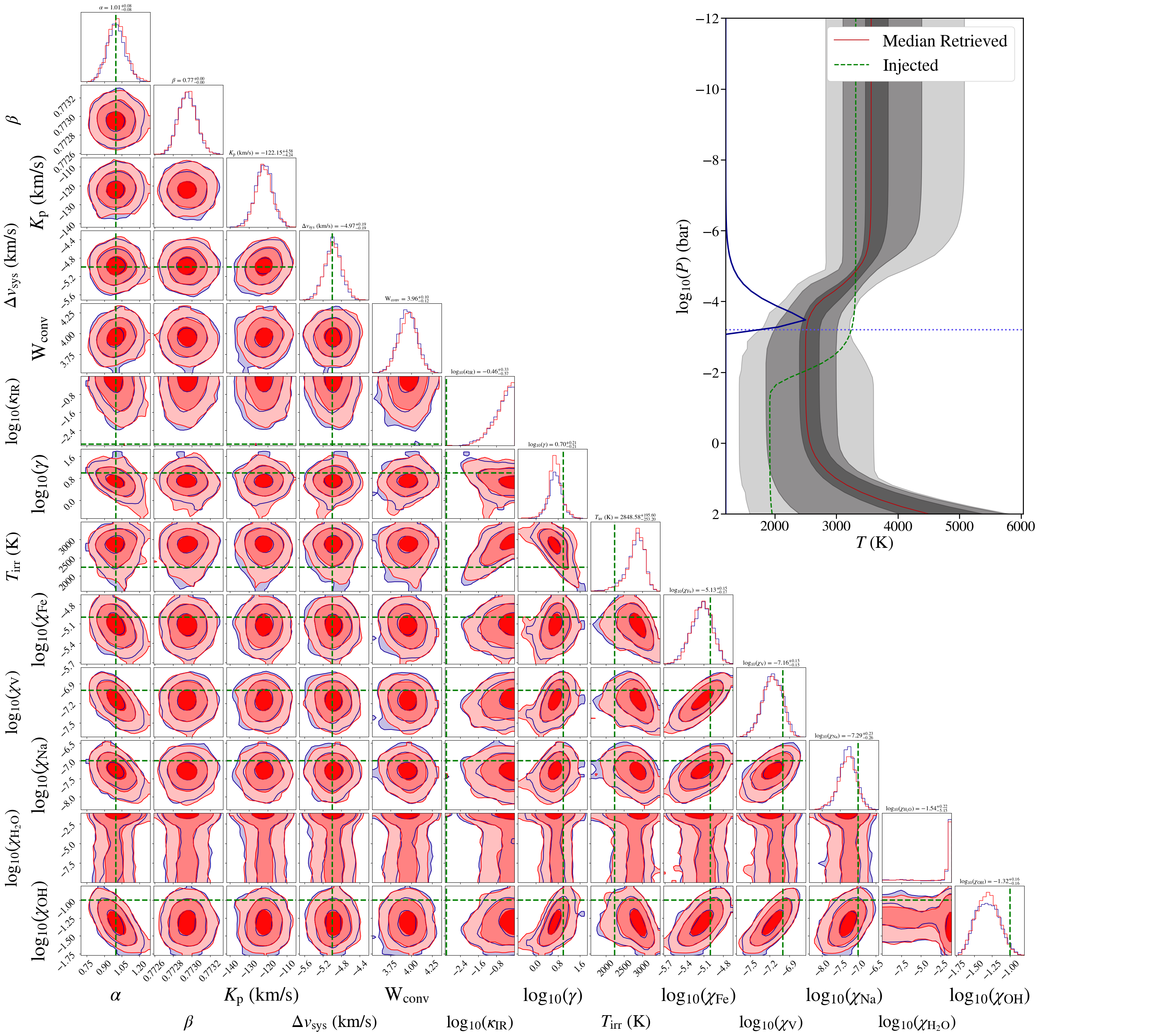}
        \caption{Atmospheric retrieval results for P2 injected into DS3 and preprocessed with \textsc{molecfit}, with the 1D and 2D marginalised posterior distributions of each of the model parameters displayed. The red and blue posterior distributions represent independent subchains, of the same MCMC chain, both converging to similar distributions. The injected values are shown as green vertical/horizontal dashed lines. The $T$-$P$ profile shown on the right was computed from 10,000 random samples of the MCMC, where the solid red curve shows the median profile, and the grey shaded regions show the 1$\sigma$, 2$\sigma$, and 3$\sigma$ contours. The injected profile is shown as a green dashed curve. The cloud deck, located at $\log_{10}(P_{\rm cloud})$ is highlighted as a blue dotted line. Below this value, the model is truncated, and thus the model inference breaks down. The contribution curve is also shown in blue.} 
        \label{Obs3_P2_Mol_corner}
\end{figure*}
\begin{figure*}[!htbp]
    \centering
    \includegraphics[width=\textwidth]{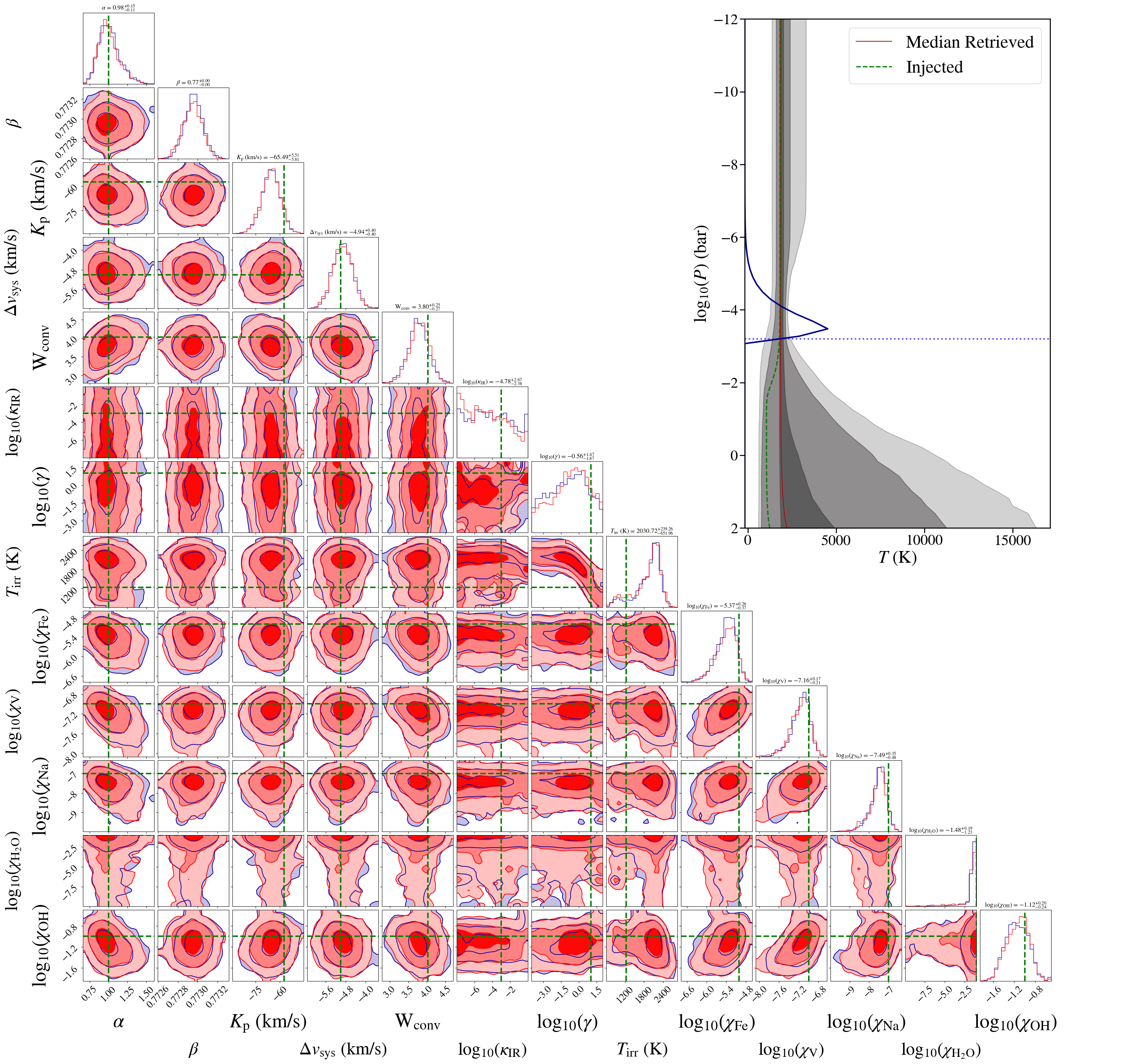}
        \caption{Similar to Fig.~\ref{Obs3_P2_Mol_corner}, for P12.} 
        \label{Obs3_P12_Mol_corner}
\end{figure*}
\end{appendix}
\end{document}